\documentclass[%
 reprint,
nofootinbib,
 amsmath,amssymb,
 aps,
]{revtex4-2}

\usepackage{amsmath}
\usepackage{mathtools}
\usepackage{graphicx}
\usepackage{dcolumn}
\usepackage{bm}
\usepackage[normalem]{ulem}
\usepackage[com pat=1.1.0]{tikz-feynman}
\usepackage{tikz}
\usepackage{comment}
\usetikzlibrary{arrows}
\usepackage[colorlinks=true, allcolors=blue]{hyperref}
\usepackage{aas_macros}
\usetikzlibrary{external}
\begin{document}

\preprint{APS/123-QED}

\title{Flavor conversions with energy-dependent neutrino emission and absorption}

\author{Chinami Kato}
 \affiliation{Faculty of Science and Technology, Tokyo University of Science, 2641 Yamazaki, Noda-shi, Chiba 278-8510, Japan}
 \email{ckato@rs.tus.ac.jp}
\author{Hiroki Nagakura}%
\affiliation{%
 National Astronomical Observatory of Japan, 2-21-1 Osawa, Mitaka, Tokyo 181-8588, Japan}%
 \author{Masamichi Zaizen}
 \affiliation{Faculty of Science and Engineering, Waseda University, Tokyo 169-8555, Japan}

\date{\today}

\begin{abstract}
Fast neutrino flavor conversions (FFCs) and collisional flavor instabilities (CFIs) potentially affect the dynamics of core-collapse supernovae (CCSNe) and binary neutron star mergers (BNSMs).
Under the assumption of homogeneous neutrinos, we investigate effects of neutrino emission and absorption (EA) by matters through both single and multi-energy numerical simulations with physically motivated setup. 
In our models, FFCs dominate over CFIs in the early phase, while EA secularly and significantly give impacts on flavor conversions.
They facilitate angular swaps, or the full exchange between electron neutrinos ($\nu_e$) and heavy-leptonic neutrinos ($\nu_x$).
As a result, the number density of $\nu_x$ becomes more abundant than the case without EA, despite the fact that the isotropization by EA terminates the FFCs earlier.
In the later phase, the system approaches new asymptotic states characterized by EA and CFIs, in which rich energy-dependent structures also emerge.
Multi-energy effects sustain FFCs and the time evolution of the flavor conversion becomes energy dependent, which are essentially in line with effects of the isoenergetic scattering studied in our previous paper.
We also find that $\nu_x$ in the high-energy region convert into $\nu_e$ via flavor conversions and then they are absorbed through charged current reactions, exhibiting the possibility of new path of heating matters.
\end{abstract}

\maketitle

\section{Introduction}
Decades of theoretical research on core-collapse supernovae (CCSNe) and binary neutron star mergers (BNSMs) have shown that neutrinos play important roles in the dynamics of these explosive transient events. The influence of neutrino physics in these phenomena has been much discussed.
One of the quantum kinetic features receiving increased attention throughout the community is collective neutrino oscillations, which are neutrino flavor conversions induced by neutrino self-interactions \cite{samuel1993,sigl1993,sigl1995,sawyer2005}.
Fast neutrino flavor conversions (FFCs) \cite{sawyer2005} and collisional flavor instabilities (CFIs) \cite{johns2021} potentially affect the dynamics. 
The sensitivity of CCSN and BNSM dynamics to these flavor conversions have been recently discussed \cite{Just2022,fernandez2022,Xiong2022,Nagakura2023,Ehring2023}, although there is very little known about their nonlinear properties.

FFC can occur independently of vacuum oscillations and its dynamics is essentially energy-independent.
It has been found that electron neutrino lepton number crossings (ELN crossings) in angular distributions are a necessary and sufficient condition for FFCs \cite{morinaga2021,Dasgupta2022}.
The associated criterion, namely ELN-XLN crossing, also offers to determine asymptotic states of FFCs in nonlinear regime \cite{Nagakura2022,zaizen2022}, where XLN denotes a heavy-leptonic neutrino lepton number.
Recent studies have found that the ELN crossings are ubiquitous in CCSNe \cite{dasgupta2017,tamborra2017,dasgupta2018,abbar2019b,nagakura2019,milad2019,milad2020,morinaga2020,abbar2020,abbar2020b,glas2020,capozzi2020,abbar2021,capozzi2021,nagakura2021d,harada2022,akaho2022} and BNSMs \cite{wu2017b,wu2017,george2020,Li2021,Just2022,sherwood2022,fernandez2022,grohs2022}.
This suggests that we need to accommodate FFCs in the theoretical modeling of CCSNe and BNSMs.

Nonlinear dynamics of FFCs have been simulated by solving the neutrino quantum kinetic equation (QKE) \cite{abbar2019,bhattacharayya2021,fernandez2022,grohs2022,martin2020,sherwood2021,sherwood2021b,wu2021,zaizen2021,abbar2022,sherwood2022b,bhttacharyya2022}.
Under the assumption of homogeneous neutrinos, the dynamics can be described in analogy with pendulum motions \cite{johns2020,Ian2022}.
Although the pendulum analogy offers deeper insights of FFCs, this toy model can not be applied in the case with neutrino-matter collisions \cite{sasaki2021,Ian2022b}; indeed, the FFC dynamics becomes qualitatively different \cite{capozzi2019,sherwood2019,martin2021,shalgar2021b,sasaki2021,kato2022,hansen2022,sigl2022}.
Recent studies also performed large-scale simulations of FFCs with collisions at various levels of approximations \cite{shalgar2022,shalgar2022b,Nagakura2022,nagakura2022b,Xiong2022,Nagakura2023}.
We refer readers to \cite{tamborra2021,capozzi2022,sherwood2022c} for recent reviews.

As mentioned above, matter collisions potentially have a significant impact on the dynamics of flavor conversions.
CFI, which is a flavor conversion induced by matter collisions, has recently attracted much attention \cite{johns2021}.
One of the necessary conditions for CFIs is the difference of reaction rates between neutrinos and antineutrinos \cite{Lin2022}.
A notable feature of CFIs is that flavor conversions occur even for isotropic neutrino distributions \cite{johns2021}.
Some recent studies suggested that CFIs can ubiquitously occur in CCSN \cite{Xiong2022} and BNSM \cite{Lin2022} environments.
It is also noteworthy that the growth rate of CFIs has a resonant structure, if the number density of electron neutrinos ($\nu_e$) and their antipartners ($\bar{\nu}_e$) is nearly equal to each other \cite{Lin2022,Liu2023}. If FFCs are parasitically induced by CFIs, the flavor conversions may be further accelerated \cite{Johns2022}.

One thing we do notice is that the nonlinear interplay between flavor conversions and matter collisions remains an open question, even in the case with homogeneous neutrino assumption.
As for FFCs, the coupling with isoenergetic scattering such as nucleon scattering has been studied \cite{shalgar2021b,kato2022,hansen2022,sasaki2021}, whereas the impact of emission and absorption on flavor conversions has not yet been investigated in detail.
It should also be mentioned that, although the interplay between CFIs and emission/absorption has been already investigated \cite{johns2021,Johns2022,Lin2022,Xiong2022b,Liu2023}, there are little studies on the cases with anisotropic distributions or non-monochromatic neutrinos \cite{Johns2022,Lin2022}.
The better understanding of roles of emission/absorption on these flavor conversions can contribute valuable information to interpret results obtained from more self-consistent and global simulations of FFCs and CFIs \cite{Nagakura2023,Xiong2022}, and it may also offer new quantum kinetic features of neutrinos.

In this paper, we perform single and multi-energy dynamical simulations of flavor evolution with neutrino emission and absorption.
Neutrinos are assumed to be homogeneous space and axial-symmetric (but anisotropic) in momentum space.
It is also assumed that the system consists of two-neutrino flavor.
In this study, we pay particular attention to the energy- and angular-dependent features of flavor conversions. One of the intriguing features found in this study is that the so-called angular swap, that corresponds to the full exchange between $\nu_e$ and heavy-leptonic neutrinos ($\nu_x$) at a certain angular region, is facilitated by emission and absorption. We will discuss the possible generation mechanism.

This paper is organized as follows.
We start with the numerical setup for dynamical simulations of flavor conversions in Section~\ref{sec:setup}. 
We then move on to investigate effects of neutrino emission and absorption on flavor conversions with the energy-independent reaction rates from various perspectives in Section~\ref{sec:single}.
In Section~\ref{sec:multi}, we extend the study to the case of energy-dependent emission and absorption.
Finally, we summarize our key findings and conclusions in Section~\ref{sec:summary}.

\section{Numerical setup} \label{sec:setup}

Time evolution of neutrinos in phase space follows the quantum kinetic equations (QKEs) \cite{sigl1993,volpe2015},
\begin{eqnarray}
 i\left( \frac{\partial}{\partial t} + \vec{v}\cdot \nabla \right) \rho(\vec{x},\vec{p},t) &=& \left[\mathcal{H}, \rho(\vec{x},\vec{p},t)\right] 
 + i \mathcal{C[\rho]}, \label{eq:rho_orig}\\
 i\left( \frac{\partial}{\partial t} + \vec{v}\cdot \nabla \right) \bar{\rho}(\vec{x},\vec{p},t) &=& \left[\bar{\mathcal{H}}, \bar{\rho}(\vec{x},\vec{p},t)\right]
 + i \mathcal{\bar{C}[\bar{\rho}]}. \label{eq:rhob_orig}
\end{eqnarray}
In these expressions, $\rho$ is the density matrix for neutrinos; $\mathcal{H}$ is the Hamiltonian potential; $C$ is the collision term. The bar description indicates the quantities for antineutrinos hereafter.
We consider two flavor system comprised of electron and heavy-leptonic neutrinos. Accordingly, $\rho$ and $\bar{\rho}$ have four components,
\begin{eqnarray}
\rho = \begin{pmatrix}
\rho_{ee} &\rho_{ex} \\
\rho_{ex}^\ast & \rho_{xx}
\end{pmatrix},
\bar{\rho} = \begin{pmatrix}
\bar{\rho}_{ee} &\bar{\rho}_{ex} \\
\bar{\rho}_{ex}^\ast &\bar{\rho}_{xx}
\end{pmatrix}.
\end{eqnarray}
It is also assumed that the neutrino distribution is spacial homogeneous and axial asymmetry in momentum space.
In this study, we include neutrino emission and absorption via charged-current interactions by surrounding matters.
Since heavy-leptons are unlikely to appear in CCSNe and BNSMs (but see \cite{Bollig:2017lki}), we ignore the charged-current reactions for $\nu_x$ and their anti-partners ($\bar{\nu}_x$).

\begin{figure}
\centering
\includegraphics[width=\columnwidth]{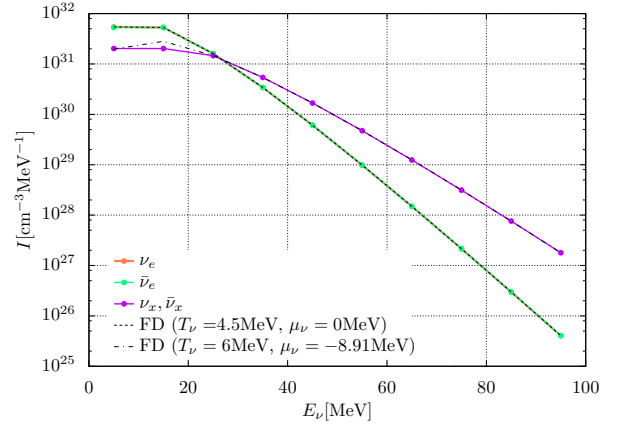}
\caption{Initial neutrino energy spectrum. Colors distinguish neutrino flavors. We also show the Fermi-Dirac distribution in black dotted ($T_\nu = 4.5$~MeV, $\mu_\nu=0$~MeV) and dot-dashed lines ($T_\nu = 6$~MeV, $\mu_\nu=-8.91$~MeV).}
\label{fig:initial_energy}
\end{figure}

\begin{figure}[htbp]
\centering
\includegraphics[width=\columnwidth]{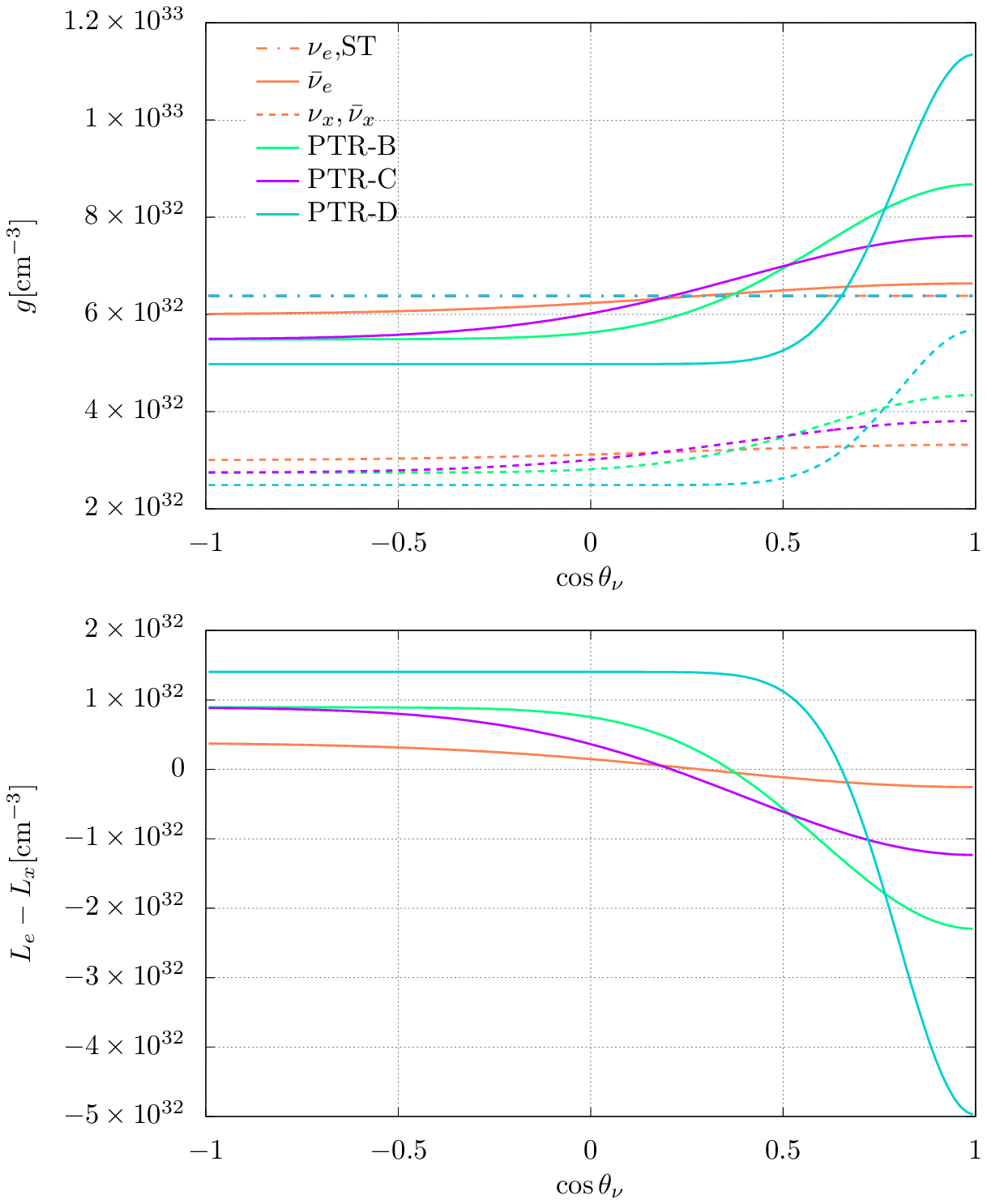}
\caption{Initial angular distributions (top) and ELN-XLN distributions (bottom). Colors and line types distinguish models and neutrino flavors, respectively. }
\label{fig:initial_angle}
\end{figure}

Under these assumptions, the QKEs can be rewritten as
\begin{eqnarray}
  && i \frac{\partial \rho_a} {\partial t} = \left[\mathcal{H}_{\nu\nu}, \rho_a\right] \nonumber \\
  && + i \begin{pmatrix}
  2\pi R_e-\left[R_e+R_a\right]\rho_{ee,a} & -\frac{1}{2}\left[R_e+R_a\right]\rho_{ex,a} \\
  -\frac{1}{2}\left[R_e+R_a\right]\rho_{xe,a}
  & 0
  \end{pmatrix},
  \label{eq:rho}\\
  && i \frac{\partial \bar{\rho}_a}{\partial t} = \left[\bar{\mathcal{H}}_{\nu\nu}, \bar{\rho}_a\right] \nonumber \\
  && + i \begin{pmatrix}
  2\pi \bar{R}_e-\left[\bar{R}_e+\bar{R}_a\right]\bar{\rho}_{ee,a} & -\frac{1}{2}\left[\bar{R}_e+\bar{R}_a\right]\bar{\rho}_{ex,a} \\
  -\frac{1}{2}\left[\bar{R}_e+\bar{R}_a\right]\bar{\rho}_{xe,a} & 0
  \end{pmatrix}, \ \ \ \ \
  \label{eq:rhob}
\end{eqnarray}
with the azimuthal-angle integrated density matrix $\rho_a \equiv \rho_a(E_\nu,\cos{\theta_\nu},t) = \int d\phi_\nu \rho(E_\nu,\cos{\theta_\nu},\phi_\nu,t)$, the emission and absorption rates for electron neutrinos, $R_e$ and $R_a$, and neutrino self-interaction potentials, 
\begin{eqnarray}
  \mathcal{H}_{\nu\nu} &=& \sqrt{2}G_F \int\int \frac{E_\nu^{\prime 2}dE_\nu^\prime d\cos{\theta_\nu^\prime}}{(2\pi)^2} \nonumber \\
  &\times& (1-\cos{\theta_\nu}\cos{\theta_\nu^\prime}) \nonumber \\ 
  &\times& (\rho_a(E_\nu^\prime,\cos{\theta_\nu^\prime},t)-\bar{\rho}_a^{\ast}(E_\nu^\prime,\cos{\theta_\nu^\prime},t)), \label{eq:ham} \\
  \bar{\mathcal{H}}_{\nu\nu} &=& \sqrt{2}G_F \int \int \frac{E_\nu^{\prime2}dE_\nu^\prime d\cos{\theta_\nu^\prime}}{(2\pi)^2} \nonumber \\
  &\times& (1-\cos{\theta_\nu}\cos{\theta_\nu^\prime}) \nonumber \\
  &\times& (\rho^{\ast}_a(E_\nu^\prime,\cos{\theta_\nu^\prime},t)-\bar{\rho}_a(E_\nu^\prime,\cos{\theta_\nu^\prime},t)). \label{eq:hamb}
\end{eqnarray}

\begin{table*}[htbp]
    \centering
    \begin{tabular}{cccccc}
    \hline \hline
    $\nu$ flavor & $n_\nu/[10^{33}{\rm cm^{-3}}]$ & $T_\nu$/[MeV] & $\mu_\nu$/[MeV] & $R_a(30{\rm MeV})/[{\rm cm^{-1}}]$ & $R_e(30{\rm MeV})/[{\rm cm^{-1}}]$ \\ \hline
    $\nu_e$& 1.28 (1.28) &4.5 & 0.0 & $2.0\times10^{-5}$ & $1.1\times10^{-6}$\\
    $\bar{\nu}_e$ & 1.26 (1.16) & 4.5 & 0.0 & $2.0\times10^{-6}$ & $1.1\times10^{-7}$ \\
    $\nu_x,\bar{\nu}_x$& 6.28 (5.78) & 6 & -8.9 & 0.0 & 0.0 \\
    \hline\hline
    \end{tabular}
    \caption{Numerical setups in this study. We refer to the SN simulation by \citet{Sumiyoshi:2012za} ($r\sim50$~km, 100~ms after the core bounce).
    The PTR-D model has the slightly smaller number of $\bar{\nu}_e$ and $\bar{\nu}_x$ and their number densities are shown in the parentheses.}
    \label{tab:setup}
\end{table*}

We refer to results of SN simulation by \citet{Sumiyoshi:2012za} to the numerical setup.
In detail, we focus on the situation of $\sim$50~km from the stellar center at 100~ms after the core bounce.
The detailed parameters are summarized in Table~\ref{tab:setup}.
In this situation, the chemical potentials of $\nu_e$ and their anti-partners ($\bar{\nu}_e$) are almost zero and the necessary condition for FFCs is in place.
In SN matter, the matter density decreases towards the stellar surface and neutrinos are decoupled from matters at a certain radius (neutrino sphere).
The position of neutrino sphere is generally $r_{\nu_e}>r_{\bar{\nu}_e}>r_{\nu_x}$, $r_{\bar{\nu}_x}$, although it depends on a neutrino energy. 
Here we focus on the region between the neutrino spheres of $\bar{\nu}_e$ and $\nu_x$ around average neutrino energy\footnote{Specifically, we consider the transport sphere for $\nu_x$ and $\bar{\nu}_x$.}.
On the basis of this fact, we employ the initial energy spectrum of neutrinos in Figure~\ref{fig:initial_energy}.
Here the quantity on the vertical axis is defined as $I\equiv E_\nu^2/(2\pi)^3\int\rho_a d\cos{\theta_\nu}$. 
It is assumed that $\nu_e$ and $\bar{\nu}_e$ are in the thermal equilibrium at all energies and at $E_\nu\gtrsim 20$ MeV, respectively.
The energy spectrum for $\nu_e$ and the higher energy part of $\bar{\nu}_e$ match with the FD of $T_\nu=4.5$~MeV and $\mu_\nu=0$~MeV (dotted line).
On the other hand, $\nu_x$ and $\bar{\nu}_x$ are already decoupled from matters at all neutrino energies in more inner region of the star. Therefore, $\nu_x$ and $\bar{\nu}_x$ spectra are consistent with the FD of $T_\nu=6$~MeV and $\mu_\nu=-8.91$~MeV (dot-dashed line) at high energies. 
It should be noted that the spectrum of $\nu_e$ and $\nu_x$ are crossed with each other at $E_\nu\sim25$ MeV.

We assume that $\nu_e$ are isotropic initially, while we use four anisotropic distributions for $\bar{\nu}_e$, $\nu_x$ and $\bar{\nu}_x$, following the previous papers \cite{shalgar2021b,Ian2022}.
Hereafter we call the angular distribution models from \citet{shalgar2021b} and \citet{Ian2022} as ST and PTR~models, respectively.
We define energy-integrated angular distributions as $g\equiv\int \rho_aE_\nu^2dE_\nu/(2\pi)^3$, 
\begin{widetext}
\begin{eqnarray}
 && g_{ee,0}(\cos{\theta_\nu}) = 0.5n_{\nu_e}, \nonumber \\
 && \bar{g}_{ee,0}(\cos{\theta_\nu}) = 
 \left\{
  \begin{array}{l}
    \left[0.47 + 0.05\exp{\left(-\left(\cos{\theta_\nu}-1\right)^2\right)}\right]n_{\nu_e}, {\rm (ST model)} \nonumber \\
    \left[0.45 - a + \frac{0.1}{b}\exp{ \left (\frac{-\left(\cos{\theta_\nu}-1\right)^2}{2b^2}\right)}\right]n_{\nu_e},{\rm (PTRmodels)} \nonumber
  \end{array}
 \right. \\
 && g_{xx,0}(\cos{\theta_\nu}) = \bar{g}_{xx,0}(\cos{\theta_\nu}) = 0.5 \bar{g}_{ee,0}(\cos{\theta_\nu}).
 \label{eq:initial_angle}
\end{eqnarray}
\end{widetext}
$n_{\nu_e}$ is the total number of electron neutrinos; $a$ and $b$ are angular-shape parameters in Table~\ref{tab:shape_param}.
These initial angular distributions are shown in the top panel of Figure~\ref{fig:initial_angle}.
The bottom panel shows the angular distributions of ELN-XLN, or $L_e-L_x \equiv g_{ee} - \bar{g}_{ee} - g_{xx} + \bar{g}_{xx}$.
All models have the angles with both of the positive and negative $L_e-L_x$ and hence FFCs are expected to be induced in all models.
It should be noted that isotropic or non-isotropic in each flavor is consistent with whether or not the flavor is in thermal equilibrium.

\begin{table}[htbp]
    \centering
    \begin{tabular}{ccc}
    \hline\hline
        model & a & b\\ \hline
        PTR-B & 0.02 & 0.4 \\
        PTR-C & 0.02 & 0.6 \\
        PTR-D & 0.06 & 0.2 \\
        \hline\hline
    \end{tabular}
    \caption{Parameters for initial angular distributions in the PTR~models.}
    \label{tab:shape_param}
\end{table}

The emission and absorption rates ($R_e$ and $R_a$) are also adopted by reference to the same situation in the realistic SN simulation.
For the reaction rate of $\nu_e$, we emulate the electron capture by free protons, which is the dominant process in this situation.
Specifically, we take $R_a(E_\nu)=2\times10^{-5}\ {\rm cm^{-1}} (E_\nu/30{\rm MeV})^2$ and $R_e(E_\nu)$ is determined through the detailed balance relation, or $R_e(1-\rho_{ee,{\rm eq}}) = R_a\rho_{ee,{\rm eq}}$. Here $\rho_{ee,{\rm eq}}$ is the Fermi-Dirac distribution (FD) for $\nu_e$ with $T_\nu$ and $\mu_\nu$ in Table~\ref{tab:setup}. 
For $\bar{\nu}_e$, the absorption rate is assumed to be $\bar{R}_a(E_\nu) = 0.1R_a(E_\nu)$, emulating the positron capture by free neutrons, and $\bar{R}_e(E_\nu)$ is determined through the detailed balance relation in the same manner as that for $\nu_e$.  
We neglect the emission and absorption for $\nu_x$ and $\bar{\nu}_x$ except in Section~\ref{subsec:nux}.

We solve the QKEs with Monte Carlo (MC) method \cite{kato2020}. This code can treat neutrino transport, matter collisions and neutrino flavor conversions self-consistently.
We ask readers to refer to \cite{kato2020} for more details.

\section{Monochromatic neutrinos} \label{sec:single}

To extract the essence of emission and absorption effects, we first perform dynamical simulations of flavor conversions with a single neutrino energy.
In section~\ref{subsec:EAeffects}, we investigate these effects using the PTR-B model, compared to the results in the absence of emission/absorption (PTR-Bwo).
We also discuss how initial angular distributions and the emission/absorption of $\nu_x$ and $\bar{\nu}_x$ affect the results in Sections~\ref{subsec:nux} and \ref{subsec:ang_depend}, respectively.
Through this section, it is assumed that all neutrinos have $E_\nu=$ 13~MeV, which is the typical average energy of $\nu_e$ in CCSNe. 
The total numbers of neutrinos for each flavor are shown in Table~\ref{tab:setup} and we adjust the energy bin width so that the initial $\nu_e$ distribution function matches the thermal equilibrium value.
The emission and absorption rates for $\nu_e$ and $\bar{\nu}_e$ are set to be the values at $E_\nu=$ 13~MeV and they satisfy the detailed-balance relation at $T_\nu=4.5$~MeV and $\mu_\nu=0$~MeV.
We consider 128 propagation directions of neutrinos as a reference resolution, following the angular distributions in Eq.~\ref{eq:initial_angle}.

\begin{table}[htbp]
    \centering
    \begin{tabular}{ccccc}
        \hline\hline
        & \multicolumn{2}{c}{$t=0$~s} &  \multicolumn{2}{c}{$t=8\times10^{-6}$~s} \\
        model & $\omega_p/[{\rm cm^{-1}}]$ & $\gamma/[{\rm cm^{-1}}]$ & $\omega_p/[{\rm cm^{-1}}]$ & $\gamma/[{\rm cm^{-1}}]$ \\ \hline
        PTR-Bwo & $0.150$ & $2.11\times10^{-2}$ &  & \\
        ST($\Gamma=0$) & $0.167$ & $4.01\times10^{-3}$ & - & - \\
        PTR-B($\Gamma=0$) & $0.150$ & $2.11\times10^{-2}$ & - & -\\        
        PTR-C($\Gamma=0$) & $0.155$ & $9.56\times10^{-3}$ & - & - \\
        PTR-D($\Gamma=0$) & $0.952$ & $0.182$ & - & - \\
        ST & 0.167 & $4.06\times10^{-3}$ & 1.55 & $7.40\times10^{-8}$  \\
        PTR-B & 0.150 & $2.12\times10^{-2}$ & 2.72 & $2.12\times10^{-6}$ \\
        PTR-C & 0.155 & $9.62\times10^{-3}$ & 2.36 & $2.38\times10^{-6}$  \\
        PTR-D & 0.952 & 0.182 & 2.23 & $-5.20\times10^{-6}$\\
        ST(multi) & 0.167 & $4.02\times10^{-3}$ & 10.3 & $1.67\times10^{-6}$\\
        PTR-B(multi) & 0.150 & $2.12\times10^{-2}$ & 1.94 & $5.75\times10^{-6}$\\
        PTR-C(multi) & 0.155 & $9.58\times10^{-3}$ & 1.57 & $1.25\times10^{-5}$  \\
        PTR-D(multi) & 0.952 & 0.182 & 1.84 & $5.88\times10^{-5}$ \\
        \hline \hline
    \end{tabular}
    \caption{Results of linear stability analysis at $t=0$~s and $t=8\times10^{-6}$~s. "-" indicates that there are only solutions with $\gamma=0$ of dispersion relations (Eq.~\ref{def:D}).}
    \label{tab:LSA}
\end{table}

\subsection{General properties of neutrino emission and absorption effects} \label{subsec:EAeffects}

\begin{figure}[htbp]
\centering
\includegraphics[width=\columnwidth]{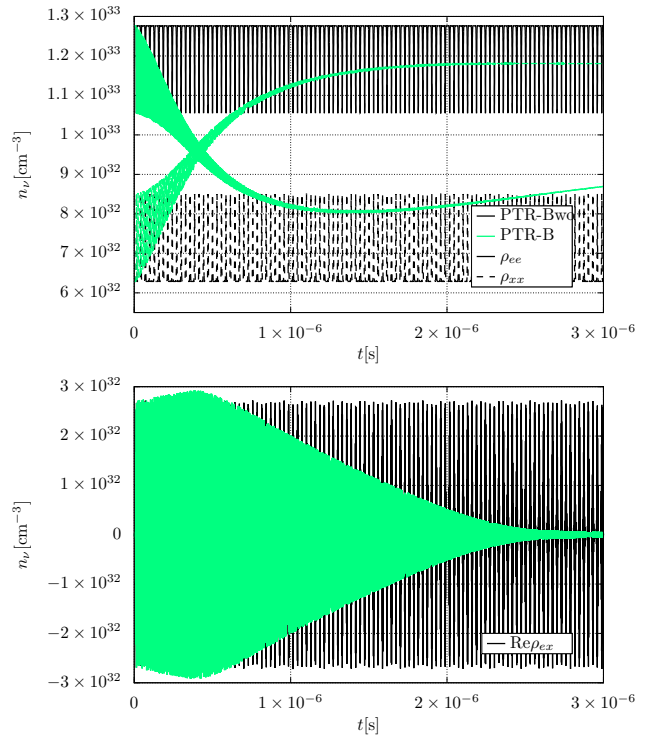}
\caption{Time evolution of number densities for $\rho_{ee}$ (top,solid), $\rho_{xx}$ (top,dotted) and Re$\rho_{ex}$ (bottom). Green and black lines denote the PTR-B and PTR-Bwo models, respectively.}
\label{fig:number_evo_noEA}
\end{figure}

\begin{figure}
    \centering
    \includegraphics[width=\columnwidth]{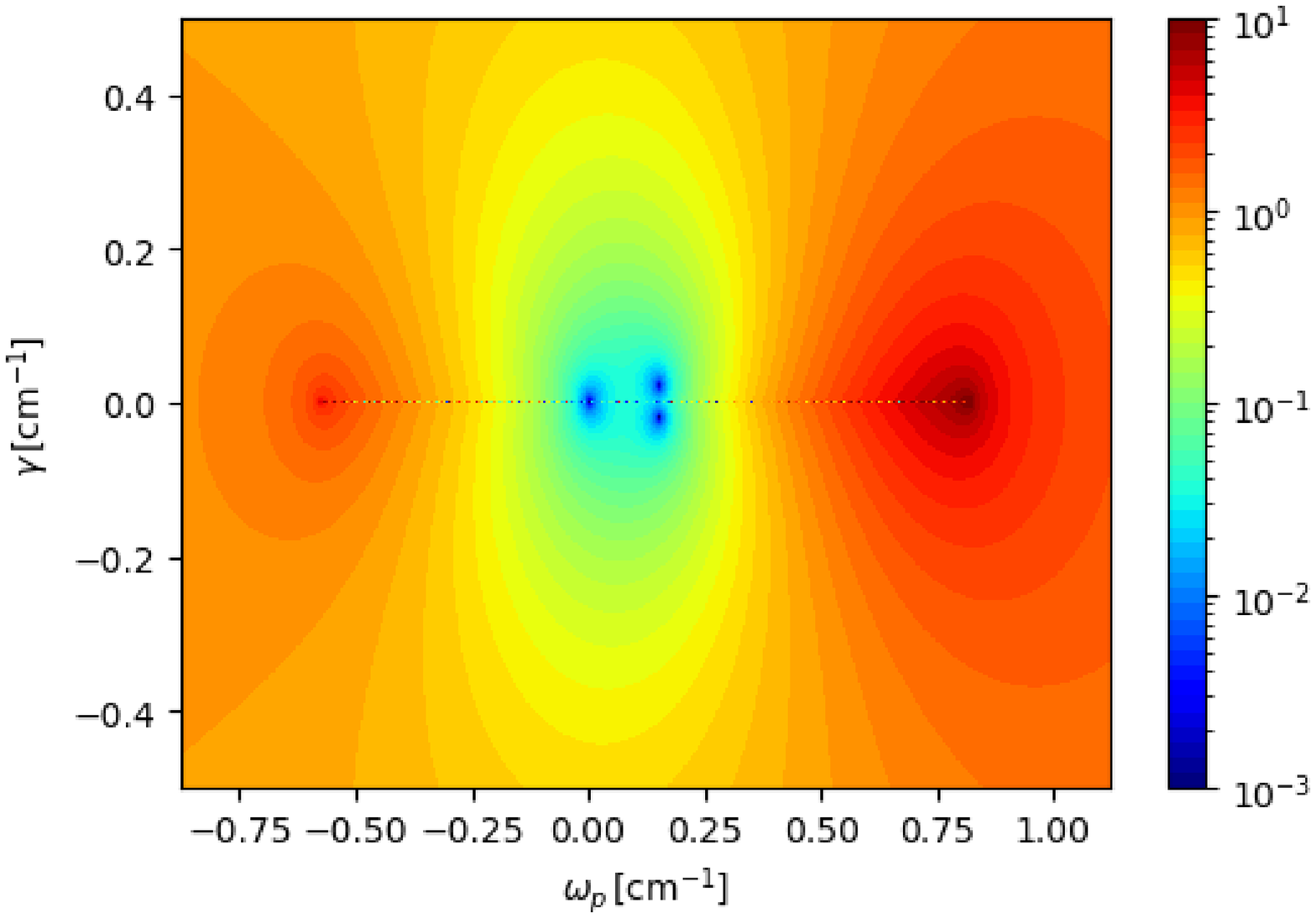}
    \includegraphics[width=\columnwidth]{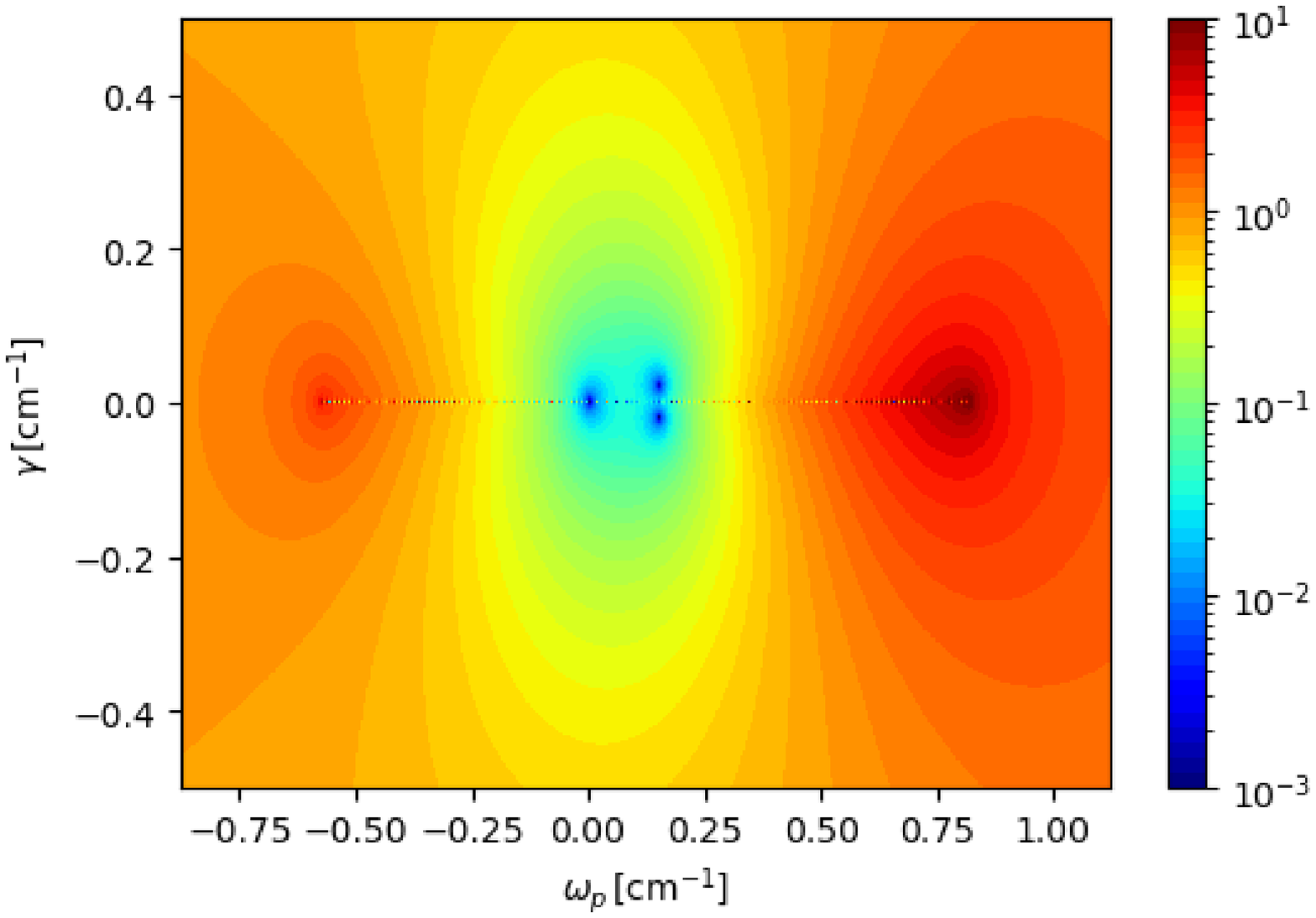}
    \includegraphics[width=\columnwidth]{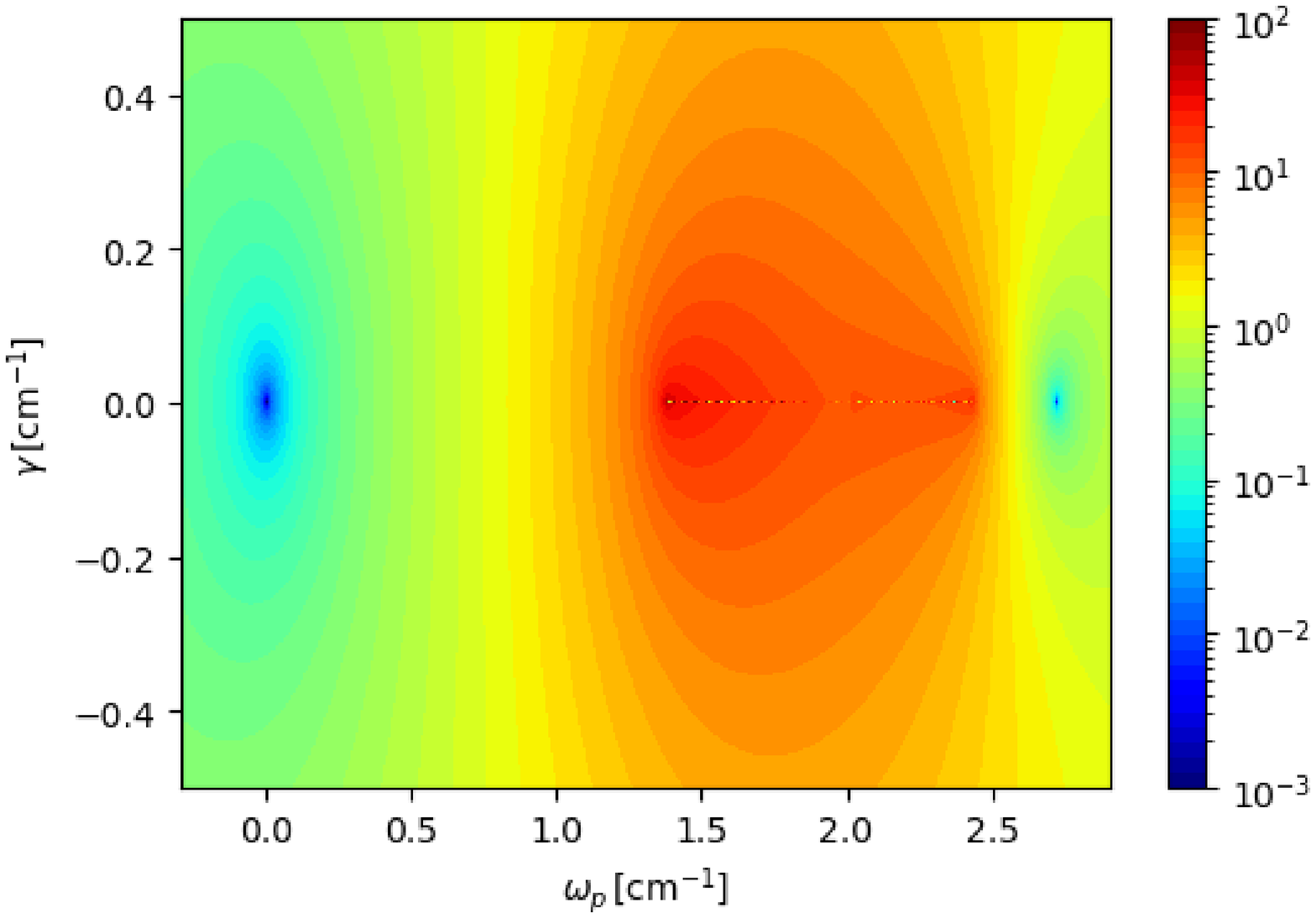}   
    \caption{Dispersion relation diagrams by the linear stability analysis. Top panel shows the result of PTR-Bwo model at $t=0$~s. Middle and bottom panels describe the results of the PTR-B model at $t=0$~s and $8\times10^{-6}$~s, respectively. }
    \label{DR_modelB}
\end{figure}

In this section, we explain effects of emission and absorption on flavor conversions using the PTR-B model.
Figure~\ref{fig:number_evo_noEA} shows the time evolution of number densities for $\rho_{ee}$ (top, solid), $\rho_{xx}$ (top, dotted) and Re$\rho_{ex}$ (bottom).
Green and black lines denote the PTR-B and PTR-Bwo models, respectively.
In the PTR-Bwo model, the number densities of all components in density matrix repeat increase and decrease periodically.
This is the typical FFC dynamics under the homogeneous assumption, which is described in analogy with pendulum motions \cite{shalgar2021b}.
In the PTR-B model, on the other hand, emission and absorption break the symmetry of the pendulum motion and the evolution of number densities is qualitatively different.
This is the same as the case of isoenergetic neutrino-matter scattering \cite{sasaki2021,Ian2022b}.
In detail, the number density of $\nu_e$ decreases until $t\sim2\times10^{-6}$~s and then it turns to increase, while that of $\nu_x$ becomes almost constant after increase.
This turning point is understood by the change of the driven mechanisms in the evolution.
From here, we look at the detailed evolutionary properties in each phase.

In the early phase ($t\lesssim2\times10^{-6}$~s), the evolution is driven by FFCs as well as the PTR-Bwo model, because the FFC timescale is much shorter than the collision one. 
Comparing with the PTR-Bwo model, however, the FFCs in the PTR-B model are more vigorous but shorter-lived.
More precisely, the FFCs are initially vigorous due to breaking the symmetry of the pendulum motion, while matter decoherence gradually attenuates the FFCs.
After the competition between these two effects, the larger number of $\nu_e$ is converted to $\nu_x$ via FFCs.
Before entering into the discussion in the nonlinear evolution, we check the results of the linear stability analysis.
We numerically solve the linearized equations for Eqs.~\ref{eq:rho} and \ref{eq:rhob} with small perturbations for off-diagonal components.
In the case without matter collisions, $\Gamma\equiv(R_e+R_a)/2$ and $\bar{\Gamma}\equiv(\bar{R}_e+\bar{R}_a)/2$ set to be 0.
The more detailed methods are described in Appendix~\ref{ap:LSA} (see also \cite{Liu2023}).
Top and middle panels of Figure~\ref{DR_modelB} show the dispersion relation diagrams for the PTR-Bwo and PTR-B models, respectively.
$\gamma$ and $\omega_p$ denote the growth rate and oscillation frequency, respectively.
Color bar describes the quantity of $|D|$ in Eq.~\ref{def:D}.
The solutions of dispersion relations correspond to $|D|\sim0$ (blue region).
Among these solutions, the unstable (stable) modes correspond to those with $\gamma>0$ ($\gamma\leqq0$).
For example, in the PTR-Bwo model, we find three blue regions and one unstable mode at $\omega_p\sim0.2~{\rm cm^{-1}}$.
The exact values of $\gamma$ and $\omega_p$ for the most unstable mode are summarized in Table~\ref{tab:LSA}.
It is found that $\gamma$ for the PTR-B model is slightly larger than that for the PTR-Bwo model, but this is a small increase for the strong flavor conversions in the nonlinear phase.
In other words, it seems difficult to predict nonlinear evolution from the linear phase study, incorporating emission and absorption. 
To see the vigor of FFCs in the nonlinear phase, the amplitudes of off-diagonal components (Re$\rho_{ex}$ and Im$\rho_{ex}$) can be a good indicator.
In the bottom panel of Figure~\ref{fig:number_evo_noEA}, the amplitude of Re$\rho_{ex}$ in the PTR-B model is larger than that in the PTR-Bwo model at $t\lesssim5\times10^{-7}$~s.
After this time, on the other hand, the amplitude decays by absorption (see the right-hand side of Eq.~\ref{eq:rho}), which indicates the suppression of FFCs.
This attenuation moves to the latter phase, or collision driven phase ($t\gtrsim 2\times10^{-6}$~s).

\begin{figure*}
\centering
\includegraphics[width=\columnwidth]{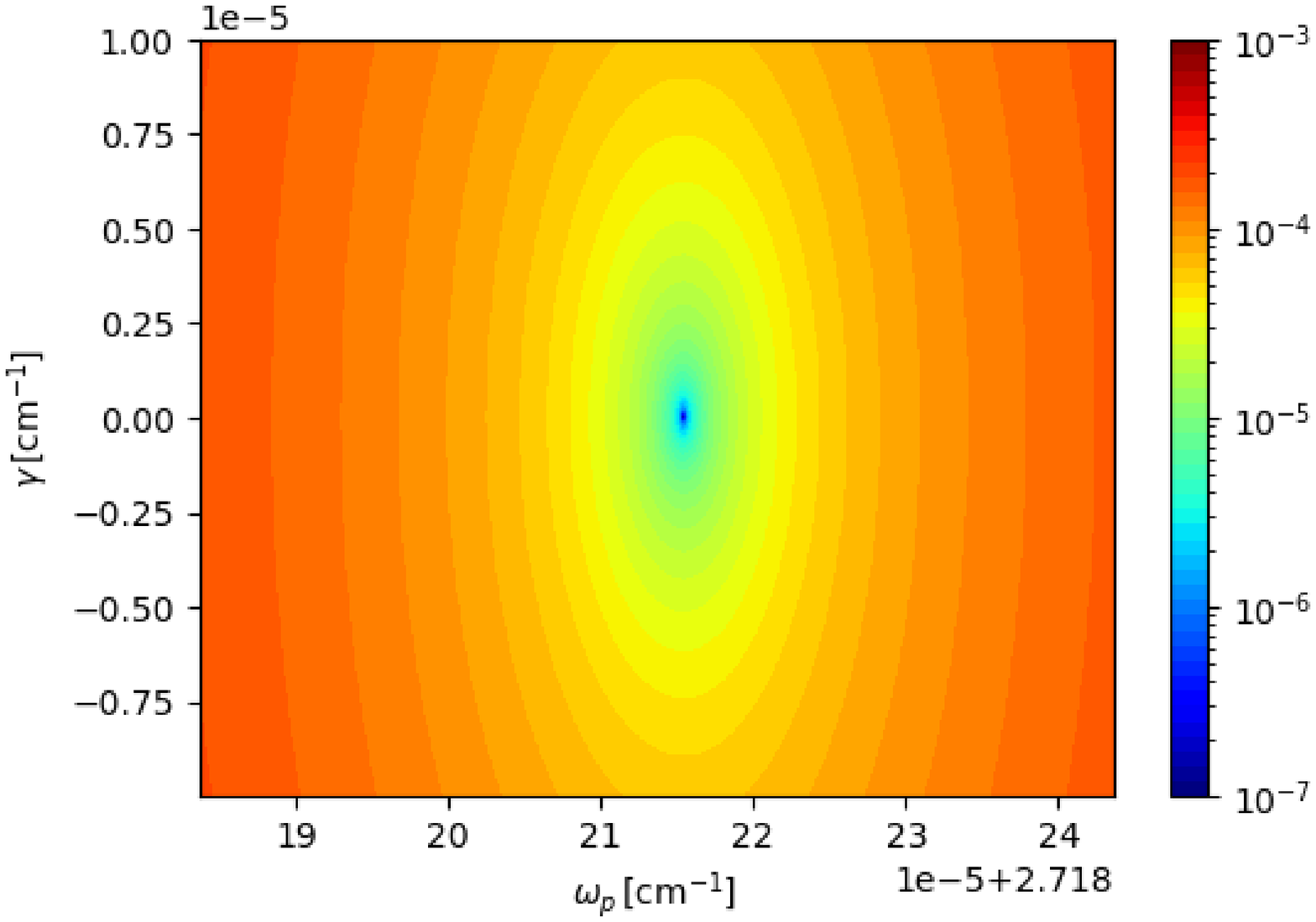}
\includegraphics[width=\columnwidth]{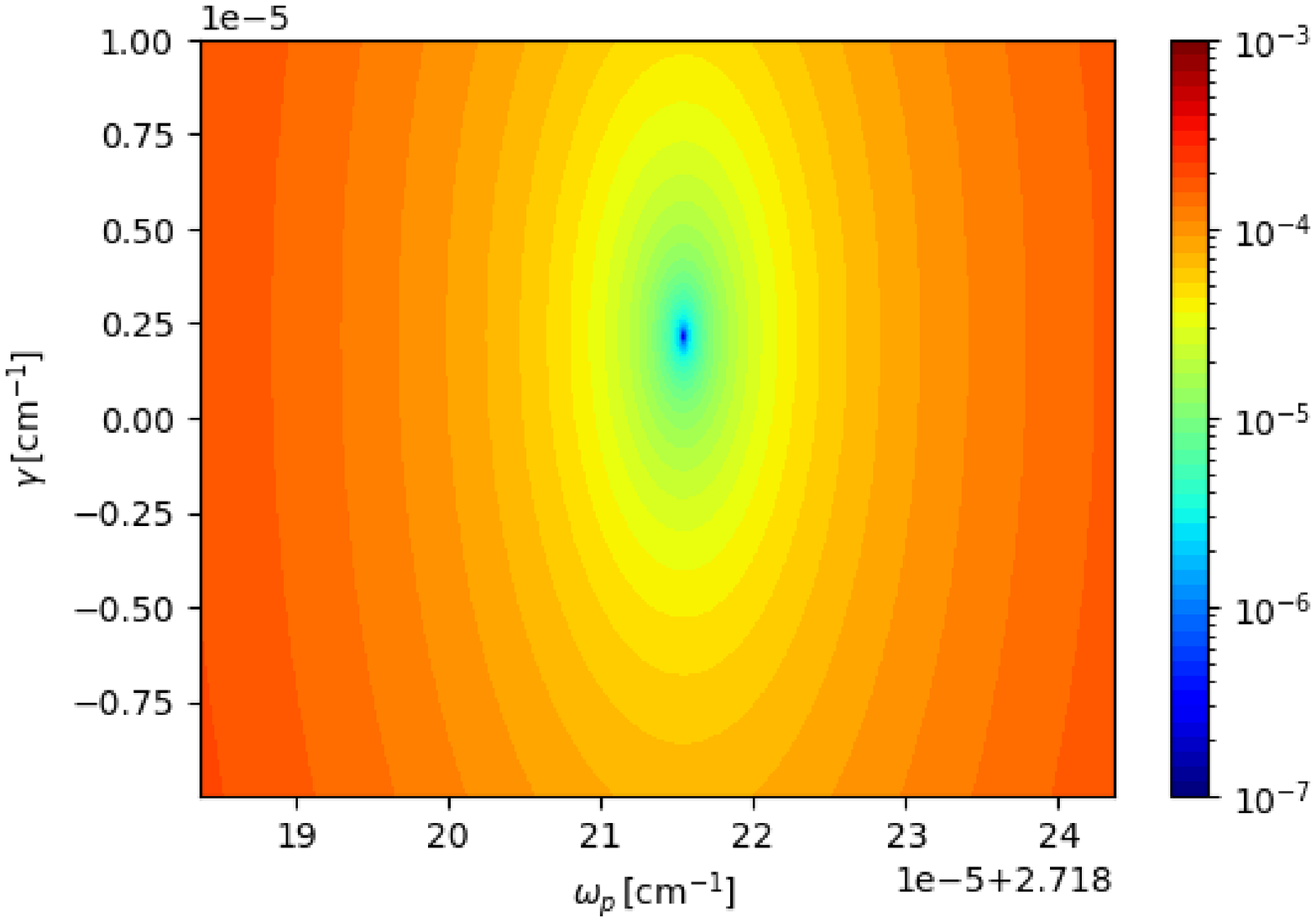}
\caption{Enlarged dispersion relation diagrams by the linear stability analysis for the PTR-B models at $t=8\times10^{-6}$~s. 
Left and right panels show the cases without and with emission and absorption, respectively.
It should be noted that we use an enlarged label in the horizontal axis. For example, the value of 22 on the axis actually corresponds to $\omega_p=2.71822$.}
\label{fig:modelB_wocol}
\end{figure*}

\begin{figure*}
\centering
\includegraphics[width=\textwidth]{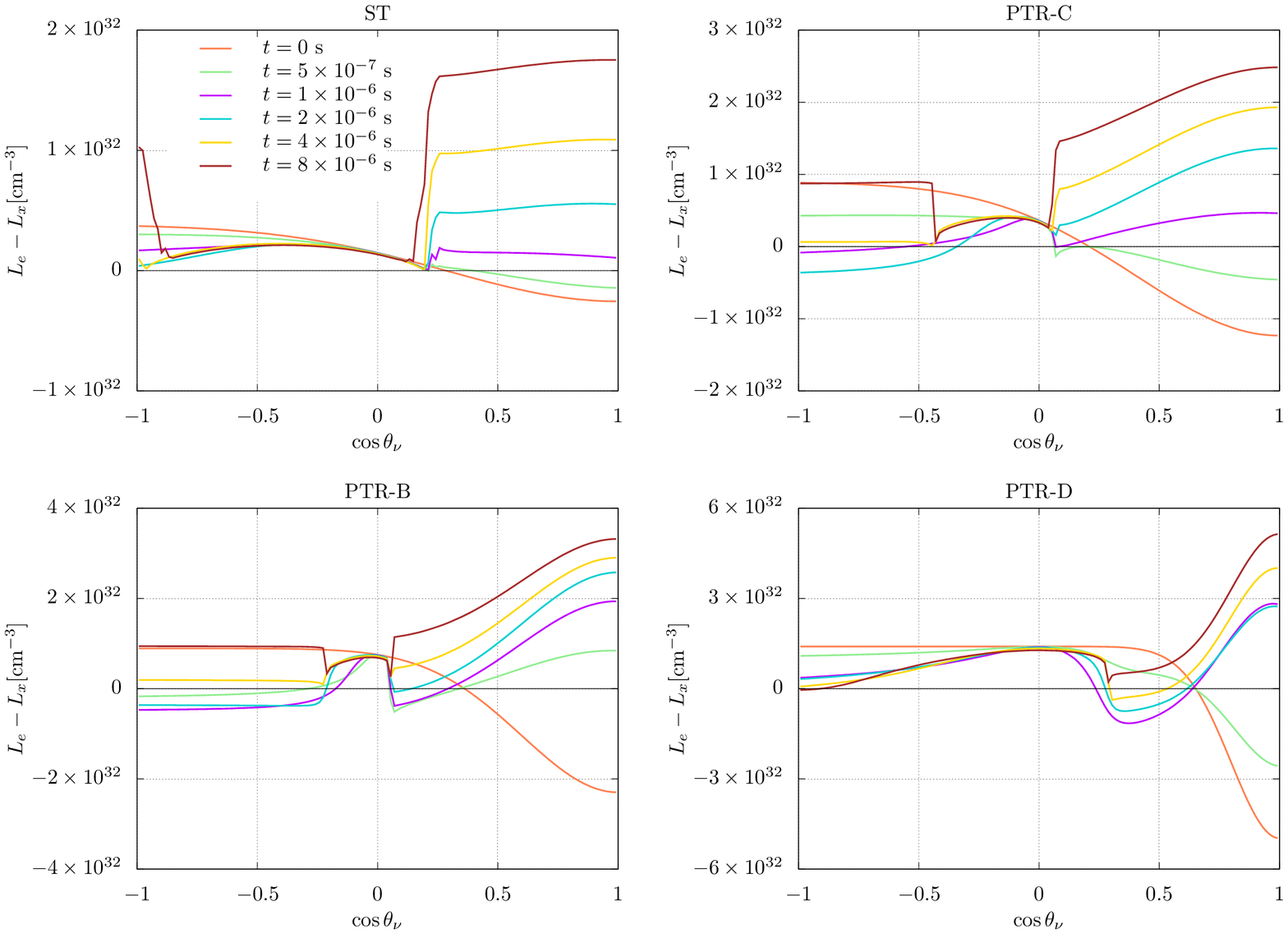}
\caption{Evolution of ELN-XLN angular distributions. Different colors denote the different time steps.}
\label{fig:LN_evo}
\end{figure*}

In the collision driven phase, CFIs and matter collisions characterize the evolution.
Since the number of $\nu_x$ is almost constant, at first glance, it appears that no flavor conversion occurs.
However, we find the unstable mode by the linear stability analysis at $t=8\times10^{-6}$~s (see Table~\ref{tab:LSA} and bottom panel of Figure~\ref{DR_modelB}).
The timescale of this mode is longer than that of interest, which is account for the constant $n_{\nu_x}$.
To deepen the understanding, we also see the growth rate for the case of $\Gamma=0$, employing the results of the PTR-B model [PTR-B($\Gamma=0$)].
It should be noted that this model is different from the PTR-Bwo one. In the linear stability analysis, both models solve Eq.~\ref{def:D} with $\Gamma=0$, but the former uses the neutrino distributions obtained from numerical simulations with emission/absorption, while the latter uses the results without them.
Figure~\ref{fig:modelB_wocol} shows the enlarged dispersion relation diagrams without (left) and with (right) emission and absorption. 
In both cases, we have the solutions at $\omega_p\sim2.718215$, but $\gamma$ is almost 0 in the collisonless case, whereas $\gamma$ is positive if collisions are taken into account. The exact values are summarized in Table~\ref{tab:LSA}. This suggests that the unstable mode is a CFI.
On the other hand, the number density of $\nu_e$ increases by emission until it will reach the thermal equilibrium with matters, or the initial state.

\begin{figure*}
\centering
\includegraphics[width=\textwidth]{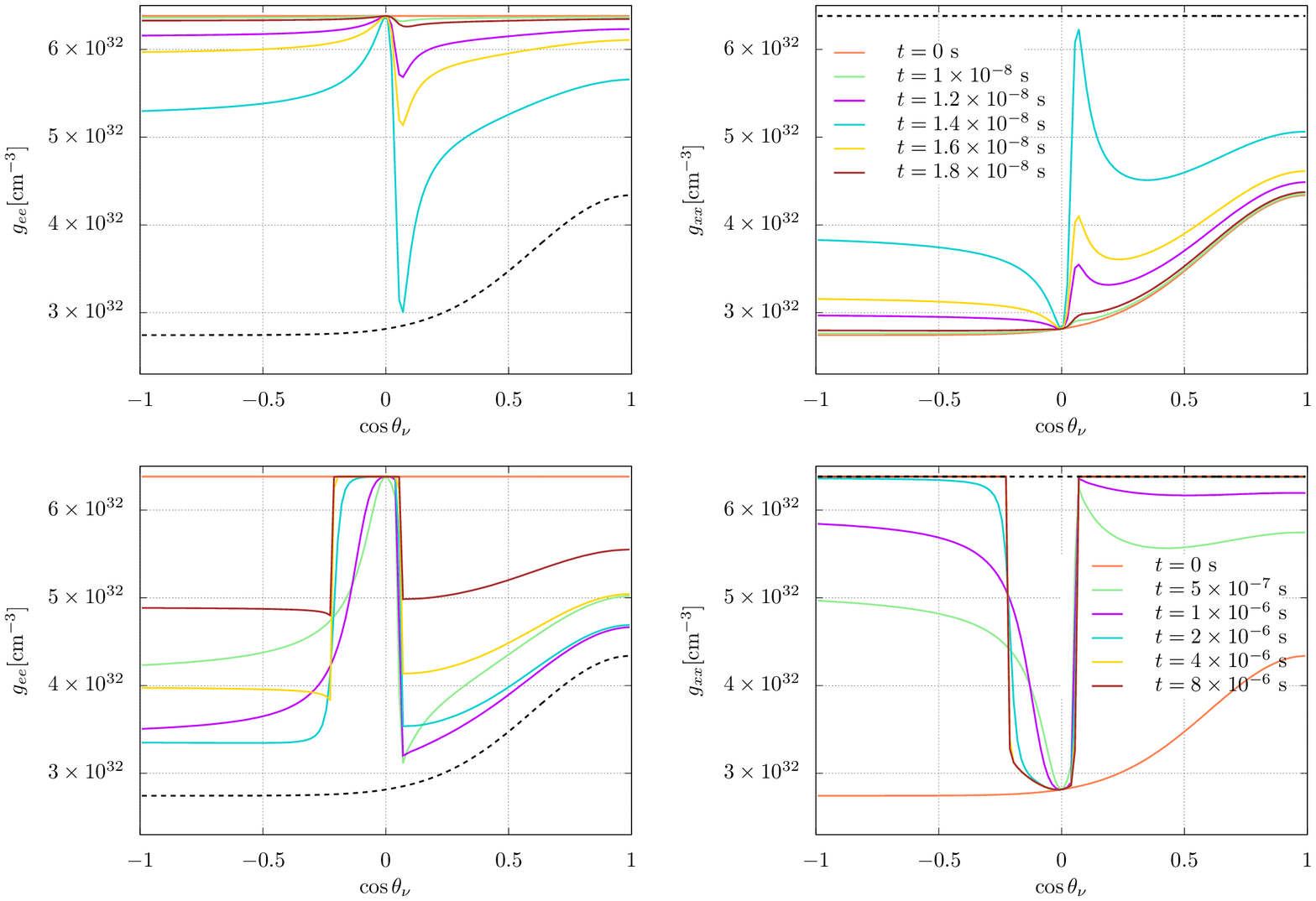}
\caption{Evolution of angular distributions in the PTR-Bwo (top) and PTR-B models (bottom). Left and right panels show results of $\nu_e$ and $\nu_x$, respectively. Different colors denote the different time steps. Each other's initial distributions are also plotted in black-dotted lines.}
\label{fig:angle_evo}
\end{figure*}

These features are also confirmed by the evolution of angular distributions.
The left bottom panel of Figure~\ref{fig:LN_evo} shows the evolution of $L_e-L_x$ angular distributions in the PTR-B model.
Different colors denote the different time steps.
It is found that the crossing point moves toward $\cos{\theta_\nu}=0$ and disappears between $2\times10^{-6}~{\rm s}<t<4\times10^{-6}$~s (cyan and yellow lines); this implies that FFCs are terminated during this time, which is consistent with the stability condition by the linear analysis (see left panel of Figure~\ref{fig:modelB_wocol}).
Figure~\ref{fig:angle_evo} shows the evolution of $g_{ee}$ (left) and $g_{xx}$ (right).
In the left bottom panel, we find that the angular distribution of $\nu_e$ deviates from the isotropic distribution by FFCs until $t\sim2\times10^{-6}$~s and then it turns to the isotropization by emission.
The angular distribution of $\nu_x$, on the other hand, is not so evolved after $t=2\times10^{-6}$~s (see right bottom panel) due to no emission and absorption processes.
These features are also  consistent with those in the number density evolution.

\begin{figure}
    \centering
    \includegraphics[width=\columnwidth]{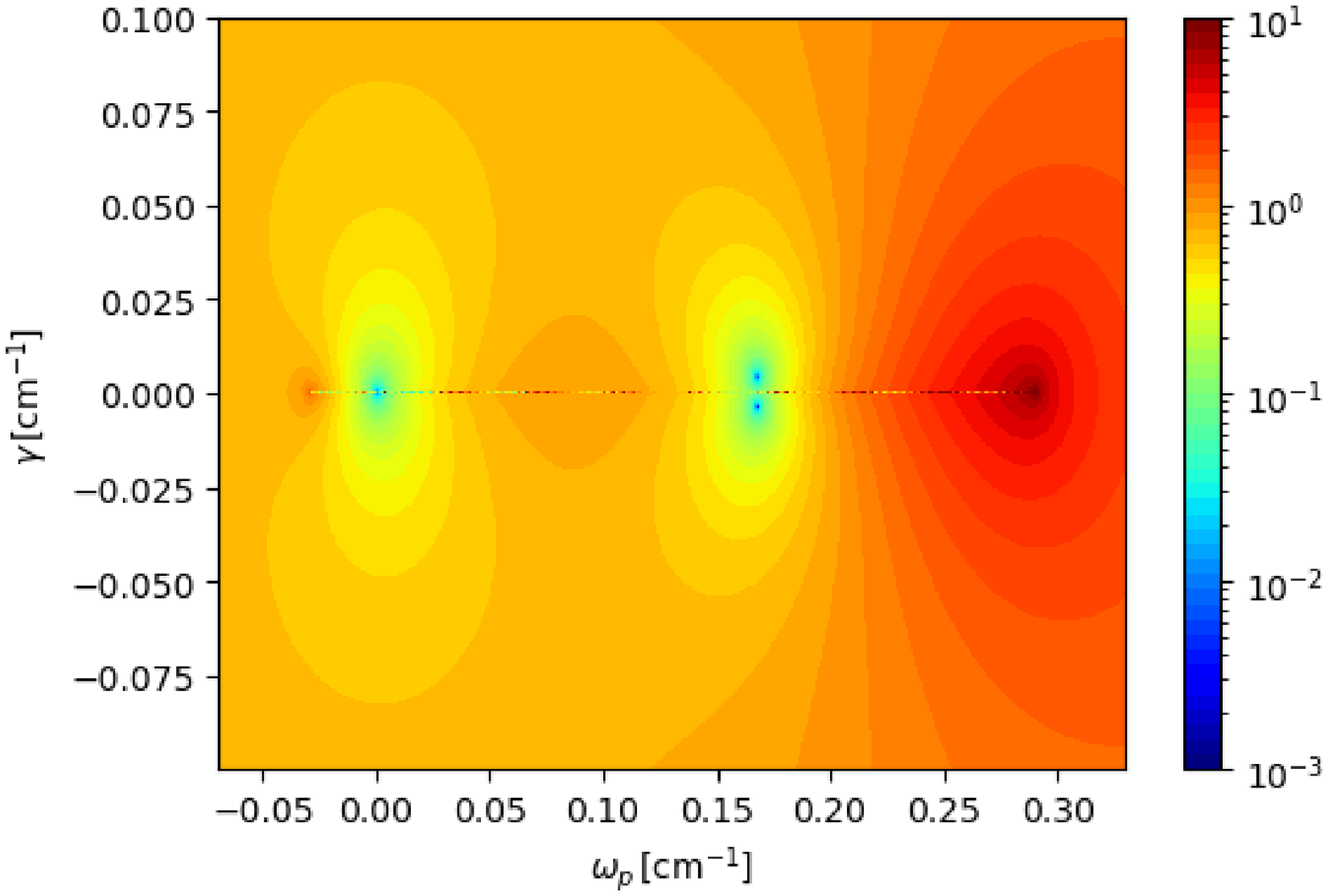}
    \includegraphics[width=\columnwidth]{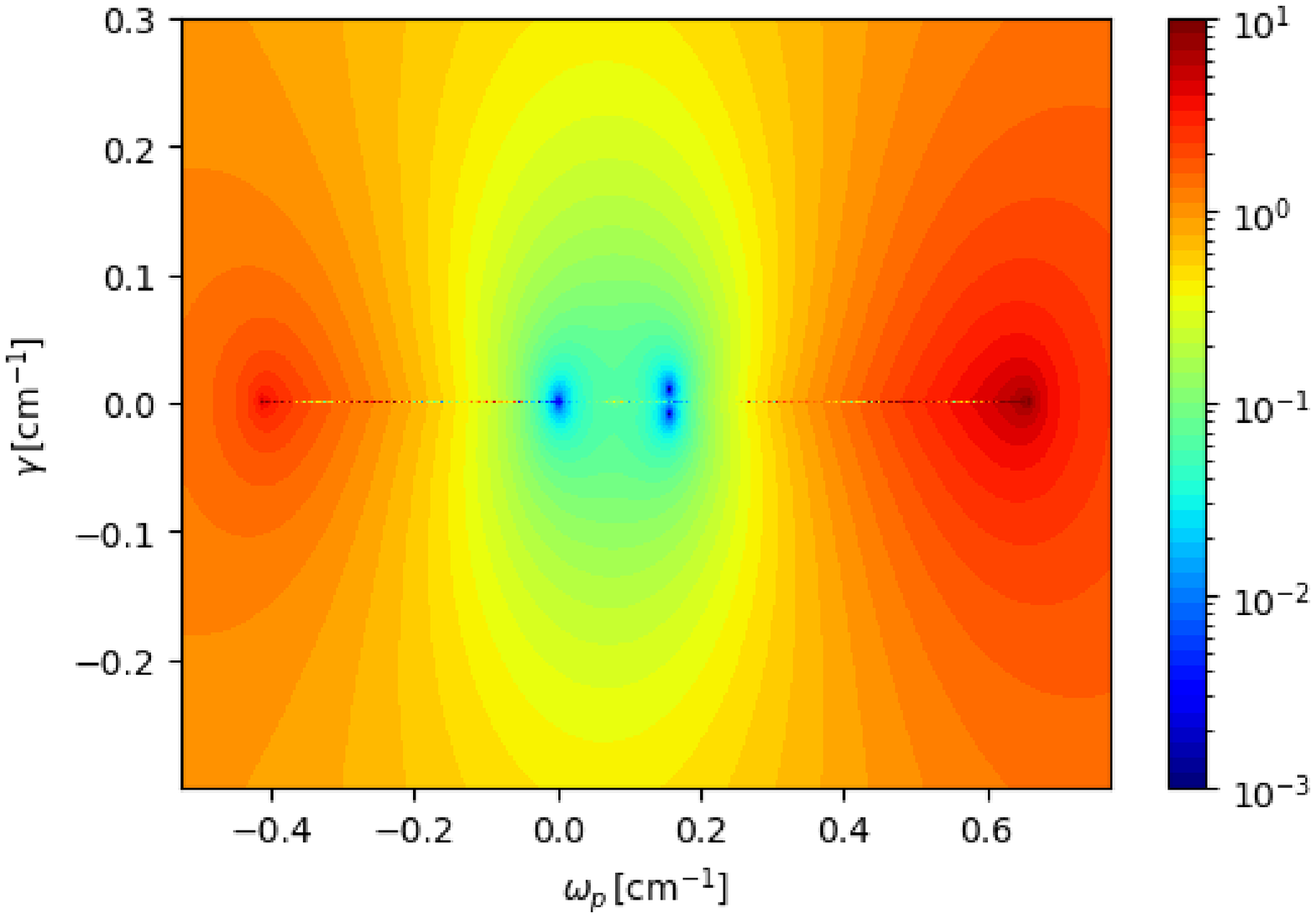}
    \includegraphics[width=\columnwidth]{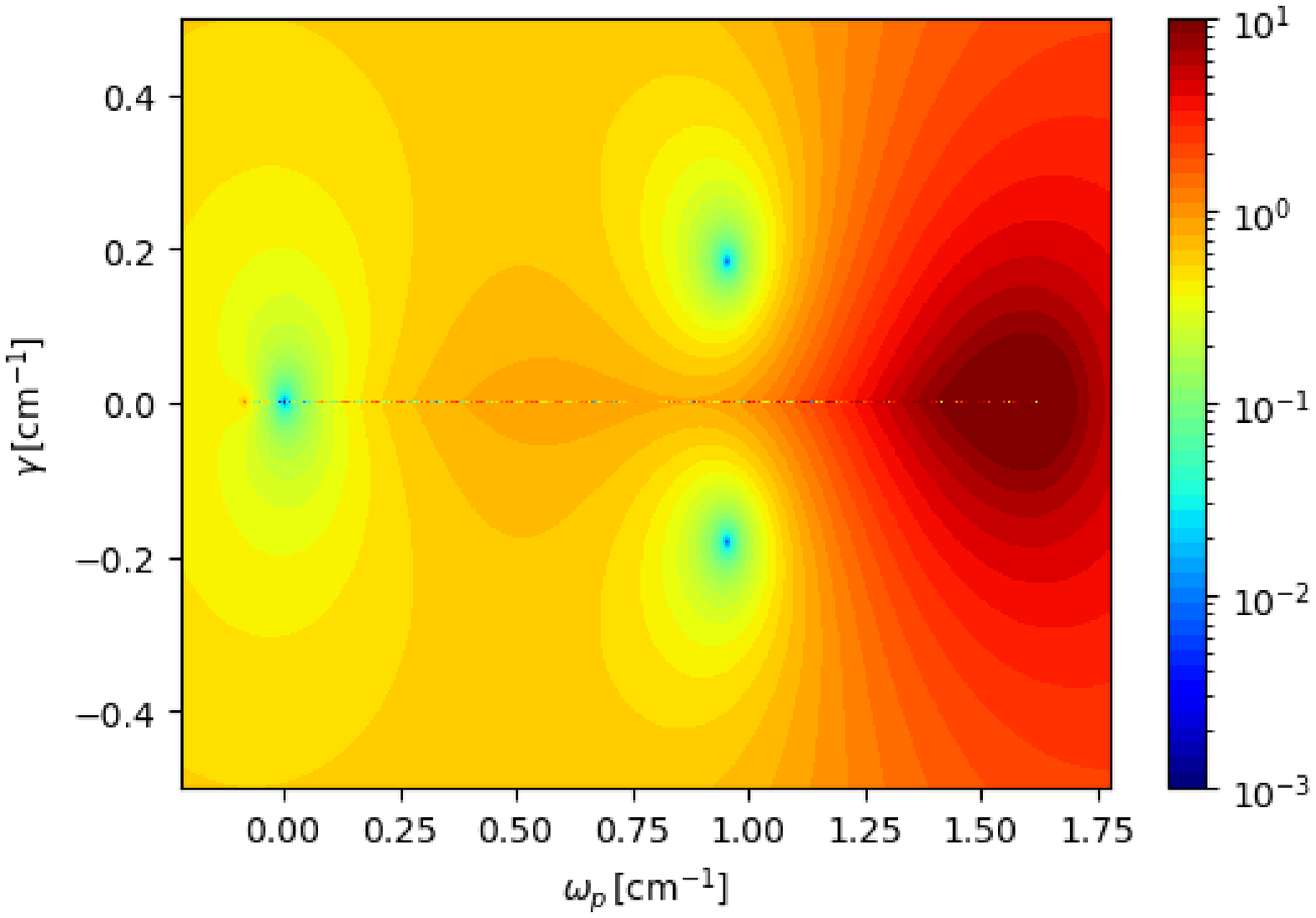}   
    \caption{Dispersion relation diagrams by the linear stability analysis at $t=0$~s. The results of ST, PTR-C and PTR-D models are shown from top to bottom panels, respectively. }
    \label{DR_initial_all}
\end{figure}

\begin{figure}
\centering
\includegraphics[width=\columnwidth]{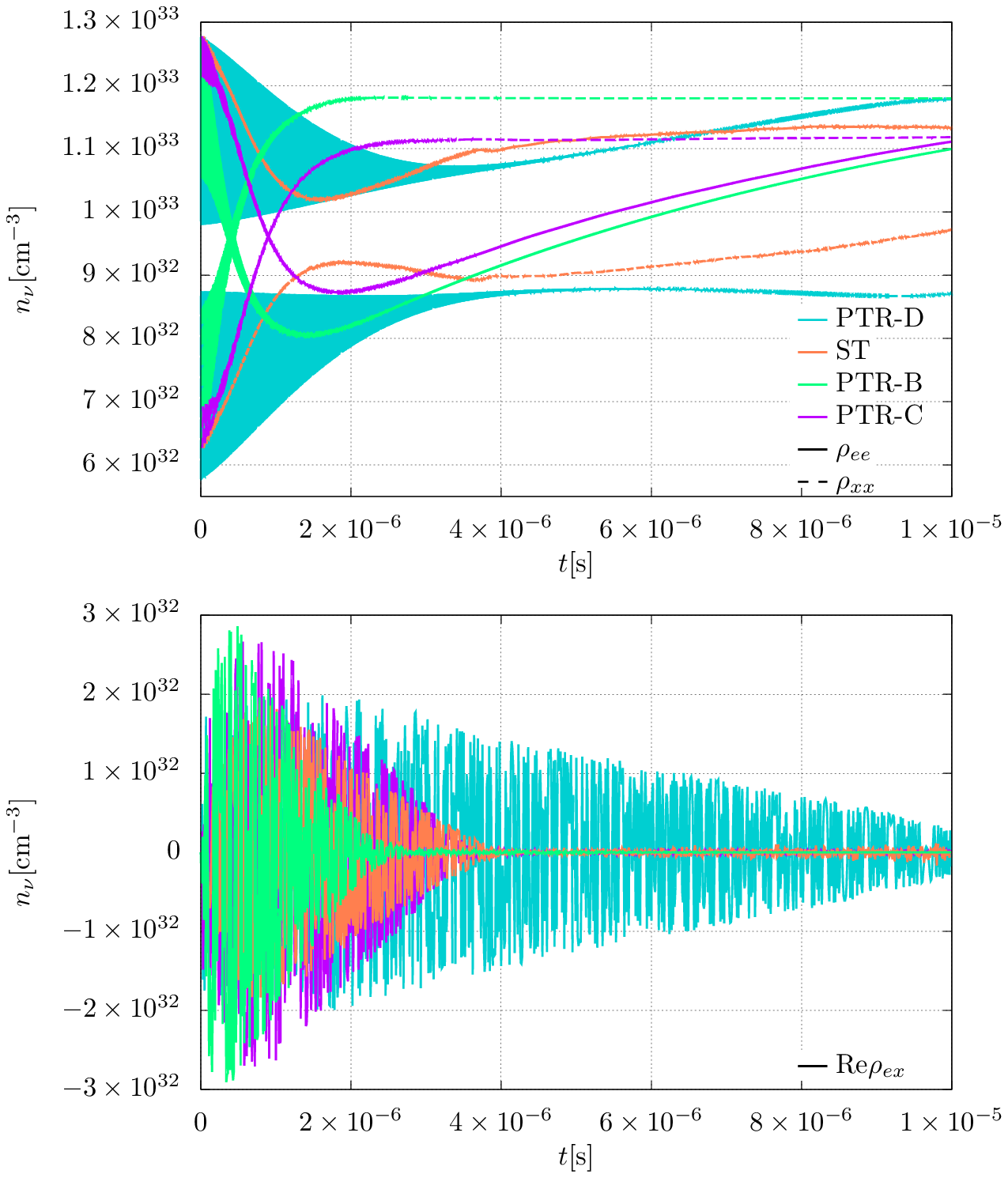}
\caption{Time evolution of number densities for $\rho_{ee}$, $\rho_{xx}$ (top) and Re$\rho_{ex}$ (bottom). Colors and line times distinguish the models and neutrino flavor, respectively. }
\label{fig:number_evo}
\end{figure}

We find another interesting feature, or an angular swap, in Figure~\ref{fig:angle_evo}.
Top panels display the results of the PTR-Bwo model.
Cyan lines denote the time step when the largest conversions occur near the $L_e-L_x$ crossing point ($\cos{\theta_\nu}\sim0.1$), and it is found that $g_{ee}$ ($g_{xx}$) at this angle is close to the initial $g_{xx}$ ($g_{ee}$).
The initial distributions are shown in black-dotted lines.
This feature is quite similar to a spectral swap observed in a slow instability \cite{duan2006}.
It is known that the total lepton number ($\rho_{ee}+\rho_{xx}$) and the quantity of $P=(\rho_{ee}-\rho_{xx})^2+4({\rm Re}\rho_{ex})^2+4({\rm Im}\rho_{ex})^2$ are conserved in each propagation direction without matter collisions \cite{hannestad2006}\footnote{$P$ corresponds to the length of polarization vector.}.
From these quantities, we derive that $g_{ee}$ and $g_{xx}$ can not change beyond each other's initial values.
Including neutrino emission and absorption, the angular swaps are found at the wider angles (see bottom panels).
Matter collisions shift
the $L_e-L_x$ crossing points slightly with time as seen in Figure~\ref{fig:LN_evo}.
On the other hand, since the collision time scale is 
much longer than that for FFCs, matter collisions can be neglected while FFCs are occurring and the similar large conversions at the new crossing points occur as the case of pure FFCs.
This is the reason for the more vigorous FFCs in the PTR-B model than those in the PTR-Bwo model.
Thus, it is worth nothing that matter collisions that are sufficiently slower than FFCs have a significant impact on the dynamics of FFCs.

\subsection{Dependence of initial angular distribution}
\label{subsec:ang_depend}

It is known that angular distributions are one of the essential factor for the dynamics of FFCs \cite{Ian2022}. Moreover, impact of matter collisions also depends on the degree of anisotropy of neutrinos.
Hence we discuss the dependence of initial angular distributions in this section.
We employ four models with different angular distributions: ST and PTR models (see Eqs.~\ref{eq:initial_angle} and Figure~\ref{fig:initial_angle}).
They have different ELN-XLN crossing points in the bottom panel of Figure~\ref{fig:initial_angle} and the order of crossing depth is PTR-D $>$ PTR-B $>$ PTR-C $>$ ST~model.
Figure~\ref{DR_initial_all} shows the dispersion relation diagrams for the ST (top), PTR-C (middle) and PTR-D (bottom) models at $t=0$~s and the $\omega_p$ and $\gamma$ of unstable modes are summarized in Table~\ref{tab:LSA}.
It is found that the growth rate is in line with the crossing depth.
This is consistent with the previous studies \cite{Ian2022}.

Figure~\ref{fig:number_evo} shows the time evolution of number densities.
The description is the same as Figure~\ref{fig:number_evo_noEA}.
We find that all models have two phases with the different driven mechanisms of evolution: FFC and matter collision driven phases, but the evolution features in each phase are different among models.
In the FFC driven phase ($t\lesssim4\times10^{-6}$~s), the features of $n_{\nu_x}$ evolution, i.e., the oscillation amplitude, vigor and lifetime of FFCs, depends on the initial angular distributions as expected from the results of linear stability analysis (see Table~\ref{tab:LSA}).
In particular, it is found that the numbers of $\nu_x$ produced via FFCs are not correlated with the initial growth rate.
For example, the PTR-D~model has the largest growth rate, whereas the final sate of $n_{\nu_x}$ is smallest among the models.
This is due to the non-linear interplay between FFCs and matter collisions.
From the previous discussion, it follows that the crossing depth in the initial angle distribution is also not an appropriate indicator of the nonlinear evolution.
This is consistent with the previous studies without matter collisions but with inhomogeneous neutrino background \cite{zaizen2022,nagakura2022b}.

\begin{figure}
    \centering
    \includegraphics[width=\columnwidth]{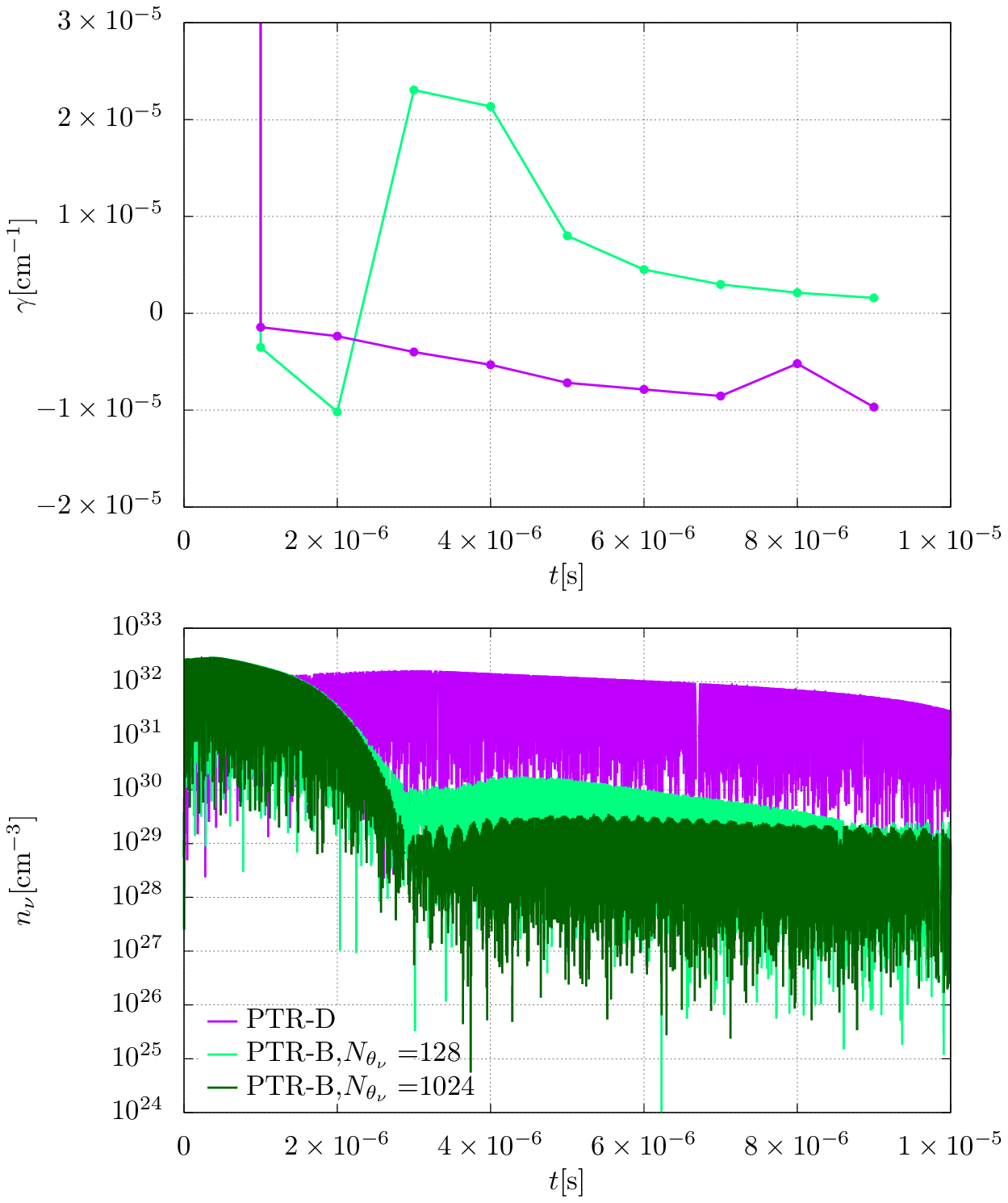}
    \caption{Time evolution of growth rate (top) and Re$\rho_{ex}$ (bottom) for the PTR-B and PTR-D models. Colors distinguish the models. It should be noted that the bottom panel uses the same data as the bottom panel of Figure~\ref{fig:number_evo}, but the vertical axis is drawn in log scale.}
    \label{fig:Relog}
\end{figure}

In the collision driven phase ($t\gtrsim4\times10^{-6}$~s), $n_{\nu_e}$'s in all models increase by emission along with the attenuation of FFCs.
On the other hand, the $\nu_x$ evolution is different among models. This shows the difference in the behavior of flavor conversions. First, we see the results of linear stability analysis. $\gamma$'s at $t=8\times10^{-6}$~s are summarized in Table~\ref{tab:LSA}. We find that the models other than the PTR-D model have unstable modes. Since these modes disappear in the collisionless cases (see the results of the $\Gamma=0$ cases in Table~\ref{tab:LSA}), they are CFIs. 
In the PTR-D model, we have only the solution with 
$\gamma<0$ and flavor conversions decay by the matter decoherence.
The matter decoherence features can be read from the evolution of Re~$\rho_{ex}$. 
To qualitatively understand the decoherence, we see the time evolution of $\gamma$ in the PTR-B and PTR-D models.
The results are shown in the top panel of Figure~\ref{fig:Relog}. In the bottom panel, we also replot the Re$\rho_{ex}$ evolution of Figure~\ref{fig:number_evo} in the logarithmic scale for the convenience.
In the former (green), at $1\times10^{-6}$~s$<t<2\times10^{-6}$~s, $\gamma$ becomes negative, which implies that the initially induced FFCs are attenuated by the matter decoherence.
After the attenuation of FFCs, $\gamma$ comes back to positive because of the CFIs.
These trends of $\gamma$ evolution match with those of Re$\rho_{ex}$ in the bottom panel including the timescale of increase/decrease.
In the PTR-D model, once $\gamma$ becomes negative at $t=1\times10^{-6}$~s, it stays throughout the simulation time. That indicates that the matter decoherence continues in the later phase. We find that this is again consistent with the Re$\rho_{ex}$ evolution.
Thus, the $\gamma$ evolution can be a useful tool to capture the evolutional trend of off-diagonal components.

The evolution of $L_e-L_x$ distributions is shown in Figure~\ref{fig:LN_evo}.
They also suggest the same features as those in the number density evolution of each model.
In the PTR-C models (right top), $L_e-L_x$'s are positive in all directions at $t\gtrsim4\times10^{-6}$~s (yellow).
This is consistent with the saturation time scale in the $n_{\nu_x}$ evolution.
Moreover, FFCs no longer occur after this time as well as the linear analysis suggests.
In the PTR-D model (right bottom), although the negative $L_e-L_x$ is found at $\cos{\theta_\nu}\sim-1$ even at $t=8\times10^{-6}$~s, the crossing is so small that the matter docoherence dominates over flavor conversions.
In the ST model, although $L_e-L_x$'s are positive at all directions after $t\sim10^{-6}$~s and the results seem to be understandable through similar considerations, we must take care of the angular resolution, as described below.

\begin{figure}
\centering
\includegraphics[width=\columnwidth]{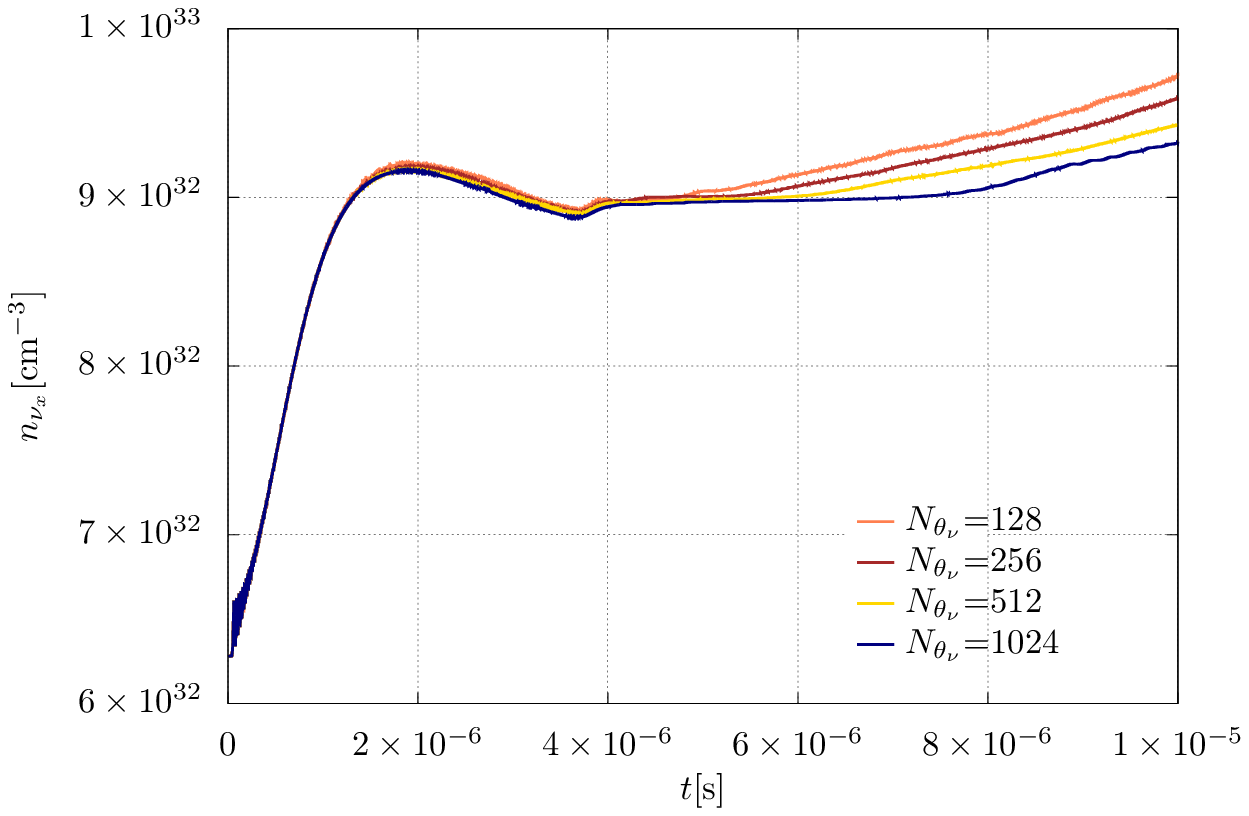}
\caption{The resolution study of $n_{\nu_x}$ evolution in the ST model. Colors distinguish the number of angular grids.}
\label{fig:modelA_reso}
\end{figure}

Finally, we should note the importance of angular resolution.
We perform the resolution study and find that the results need attention only with respect to the ST model.
Its $n_{\nu_x}$ evolution is shown in Figure~\ref{fig:modelA_reso}.
Different colors denote different number of angular grids, or $N_{\theta_\nu}$.
It is found that the increment of $n_{\nu_x}$ in the later phase ($t\gtrsim6\times10^{-6}$~s) becomes more gradual with the angular resolution.
This is because the fine structure is not accurately resolved in the case of low resolution and moreover the spurious unstable modes artificially enhance the evolution \cite{sarikas2012}.
This fact again makes us aware of the importance of the angular distribution in FFCs.
It should be noted that we confirm the reference resolution with $N_{\theta_\nu}$=128 is sufficient in the other models.

\begin{figure}
\centering
\includegraphics[width=\columnwidth]{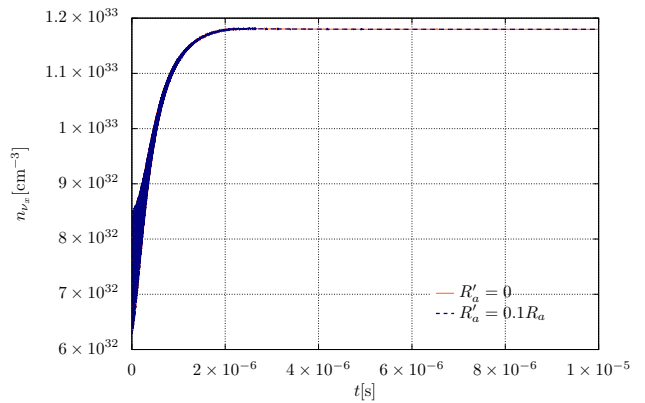}
\caption{The $n_{\nu_x}$ evolution in the cases of $R_a^\prime=0$ (coral, solid) and $0.1R_a$ (navy, dotted). We employ the PTR-B model as the initial angular distributions.}
\label{fig:nux_depend}
\end{figure}

\subsection{Emission and absorption of $\nu_x$ and $\bar{\nu}_x$} \label{subsec:nux}

$\nu_x$ and $\bar{\nu}_x$ also interact with surrounding matters via thermal processes in CCSNe and BNSMs, although their reaction rates are smaller than those of $\nu_e$ and $\bar{\nu}_e$.
If emission and absorption of $\nu_x$ and $\bar{\nu}_x$ are taken into account, they are also thermalized as well as $\nu_e$ and the evolution of $n_{\nu_x}$ may be changed.
We hence investigate effects of emission and absorption for $\nu_x$ and $\bar{\nu}_x$ in this section.
The collision term $C$, or the second term in the right hand sides of Eq.~\ref{eq:rho}, is written as
\begin{widetext}
\begin{eqnarray}
 &&C = \begin{pmatrix}
  2\pi R_e-\left[R_e+R_a\right]\rho_{ee,a} & -\frac{1}{2}\left[R_e+R_a+R^\prime_e+R^\prime_a\right]\rho_{ex,a} \\
  -\frac{1}{2}\left[R_e+R_a+R^\prime_e+R^\prime_a\right]\rho_{xe,a} &  2\pi R^\prime_e-\left[R^\prime_e+R^\prime_a\right]\rho_{xx,a}
  \end{pmatrix},
\end{eqnarray}
\end{widetext}
with the $\nu_x$ emission and absorption rates $R^\prime_e$ and $R^\prime_a$.
For antineutrinos, the same reaction rates are introduced into $\bar{C}$ in the same manner.
The absorption rates are set to be $R^\prime_a = 0.1R_a$ and the emission rates are determined by the detailed balance relation with $T_\nu = 4.5$~MeV and $\mu_\nu=0$~MeV.
The angular distributions in the PTR-B model are employed initially.
Figure~\ref{fig:nux_depend} shows the $n_{\nu_x}$ evolution.
Coral and navy lines denote the cases of $R_a^\prime=0$ and $0.1R_a$, respectively.
We find that the deviation from the case of no $\nu_x$ emission and absorption processes is $\sim5\times10^{-3}$\% at $t=10^{-5}$~s owing to the slow collision timescale ($\sim1.6\times10^{-3}$~s).
It is confirmed that thermalization effects for $\nu_x$/$\bar{\nu}_x$ are negligible unless their reactions become comparable to $\nu_e$ and $\bar{\nu}_e$.

\section{Non-monochromatic neutrinos} \label{sec:multi}

In CCSNe and BNSMs, neutrinos have non-monochromatic energy spectrum and the reaction rates of neutrino-matter interactions are energy dependent.
Hence it is natural to consider the evolution of neutrino flavors with matter collisions in the multi-energetic treatment.
In this section, we employ the neutrino energy spectrum and the energy-dependent reaction rates into numerical simulations.
The initial energy spectrum are described in Figure~\ref{fig:initial_energy} and they are consistent with the realistic CCSN simulation.
We adopt 10 energy meshes in the range of $0-100$~MeV.
As the single-energy case, the energy-integrated angular distributions follow Eqs.~\ref{eq:initial_angle} of four models (see also Figure~\ref{fig:initial_angle}) and the same number of neutrinos are employed in Table~\ref{tab:setup}.
We distribute 1,280 MC samples in a neutrino phase space.

\begin{figure}
\centering
\includegraphics[width=\columnwidth]{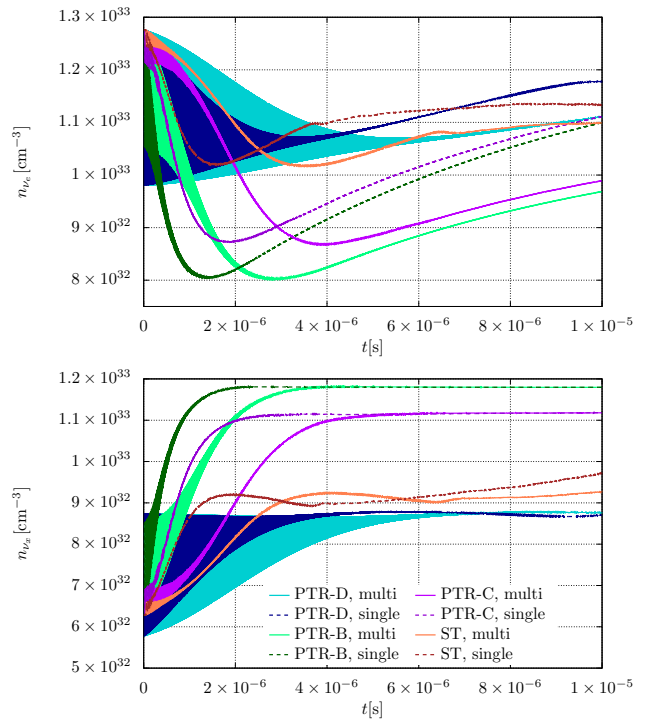}
\caption{Time evolution of $n_{\nu_e}$ (top) and $n_{\nu_x}$ (bottom) in the case of multi-energy. Colors distinguish the angular models and the results of single-energy cases are also plotted in darker dotted lines.}
\label{fig:number_evo_multi}
\end{figure}

\begin{figure*}
\centering
\includegraphics[width=\textwidth]{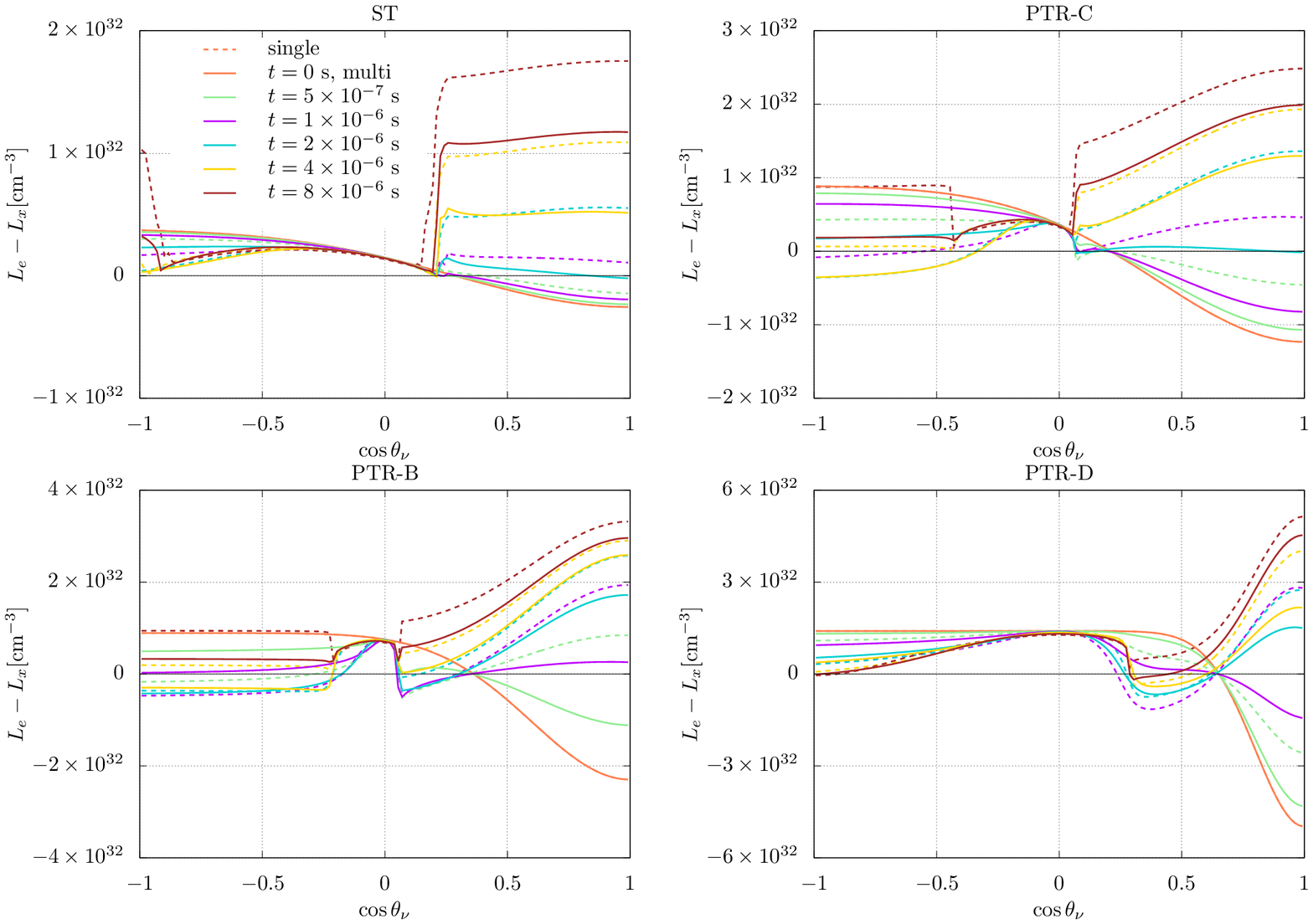}
\caption{The same figures as Figure~\ref{fig:LN_evo} but for the multi-energy case.}
\label{fig:LN_evo_multi}
\end{figure*}

Figure~\ref{fig:number_evo_multi} shows the time evolution of $n_{\nu_e}$ (top) and $n_{\nu_x}$ (bottom).
Colors distinguish the models, and the results of single-energy cases are also plotted in darker dotted lines.
We find that these evolution has two phases, FFC and collision driven phases, as same as the single-energy case.
First, we discuss effects of multi-energetic treatment using the PTR-B model as a reference (green line).
At $t\lesssim4\times10^{-6}$~s, FFCs drive the evolution as well as the single-energy case.
Table~\ref{tab:LSA} shows the results of linear stability analysis.
We find that the growth rate is almost the same as that in the single-energy case.
This fact is attributed to the property that the dynamics of pure FFCs is energy-independent and depends on the number densities of neutrinos (see Eq.~\ref{eq:rho}).
On the other hand, the overall flavor evolution is delayed by effects of energy-dependent collisions.
For $E_\nu\lesssim13$~MeV, the reaction rates are smaller than that in the single-energy case, and the decoherence by the absorption of off-diagonal components is weak.
As see in Figure~\ref{fig:initial_energy}, 
the fraction of neutrinos with $E_\nu\lesssim13$~MeV to the total number of neutrinos is $\sim84\%$ initially and these neutrinos contribute to the self-interaction Hamiltonian $H_{\nu\nu}$.
This dominancy of low energy neutrinos leads to the slower attenuation of FFCs, and hence the lifetime of FFCs is extended in the multi-energy treatment.
The extension of FFC lifetime is also confirmed by the evolution of $L_e-L_x$ evolution in the left bottom panel of Figure~\ref{fig:LN_evo_multi}.
The multi- and single-energy results are shown in solid and dashed lines, respectively.
At $t=4\times10^{-6}$~s, $L_e-L_x$'s are positive at all angles in the single-energy case, whereas the negative $L_e-L_x$ appears at $\cos{\theta_\nu}\lesssim-0.2$ in the multi-energy case.
This indicates that FFCs in the multi-energy case still survive until this time.
The crossings in this case disappear at $t=8\times10^{-6}$~s.

In our previous paper \cite{kato2022}, we have proposed the $\chi$ diagnostics to quantify the impact of multi-energy effects on FFCs with a $\chi$ parameter, or $\chi=\mid(\langle E_\nu\rangle-\langle RE_\nu\rangle)/(\langle E_\nu\rangle+\langle RE_\nu \rangle)\mid$. $\langle E_\nu \rangle$ and $\langle RE_\nu \rangle$ are the average energy of neutrinos and reaction-weighted one, respectively. In the monochromatic neutrino case, $\chi$ exactly becomes 0. If $\chi\sim0$ in the multi-energy case, neutrinos with a particular energy dominantly contribute to a flavor evolution and results are close to single-energy ones. As $\chi$ increases, the deviation from single-energy results becomes larger, and the multi-energy treatment is essential. We derive $\chi=0.281$ for the PTR-B model, which indicates that the multi-energy treatment is essential in this model. This is consistent with numerical results.

\begin{figure}
    \centering
    \includegraphics[width=\columnwidth]{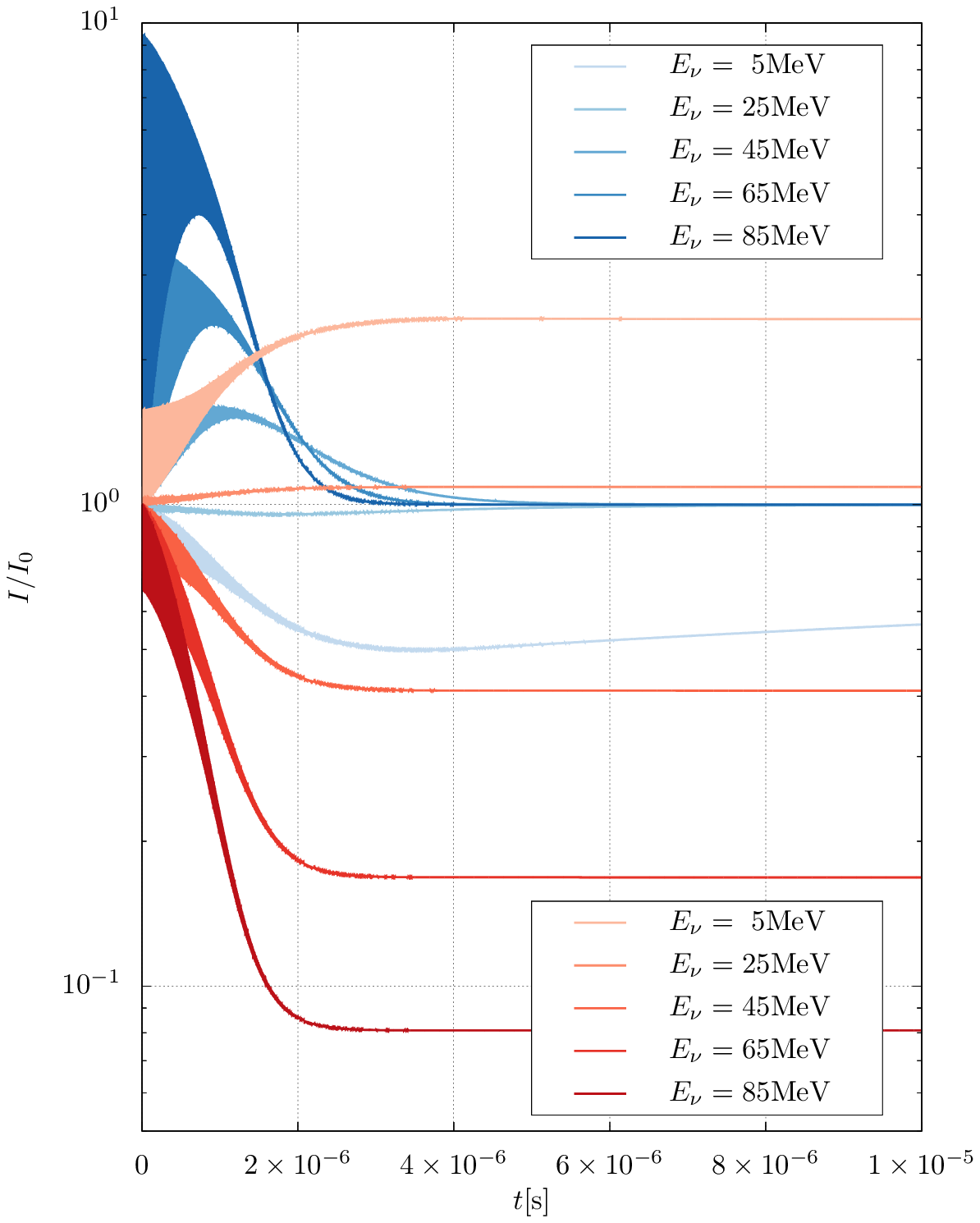}
    \caption{Time evolution of $I_{ee}$ (blue lines) and $I_{xx}$ (red lines) in the PTR-B model. We normalize $I$'s by the initial values $I_0$'s in each neutrino energy. Colors get darker with neutrino energies.}
    \label{fig:spe_evo}
\end{figure}

Similar as scatterings, the energy-dependent emission/absorption generates rich energy-dependent features in FFCs.
Figure~\ref{fig:spe_evo} shows the time evolution of $I_{ee}$ (blue) and $I_{xx}$ (red).
We normalize $I$'s by the initial values $I_0$'s in each neutrino energy.
Colors get darker with neutrino energies. 
It is found that $I_{ee}/I_{ee,0}$'s for $E_\nu\lesssim25$~MeV increase with time, whereas they decrease for $E_\nu\gtrsim25$~MeV.
This is attributed to the initial spectral crossing between $I_{ee}$ and $I_{xx}$ at $E_\nu\sim25$~MeV (see Figure~\ref{fig:initial_energy}). Since FFCs attempt to eliminate the difference between $\nu_e$ and $\nu_x$, the direction of conversions differs between high and low energies.
Related to this, we find that flavor conversions occur strongly at the higher energy because of the larger difference between $I_{ee}$ and $I_{xx}$.

\begin{figure*}
\centering
\includegraphics[width=\textwidth]{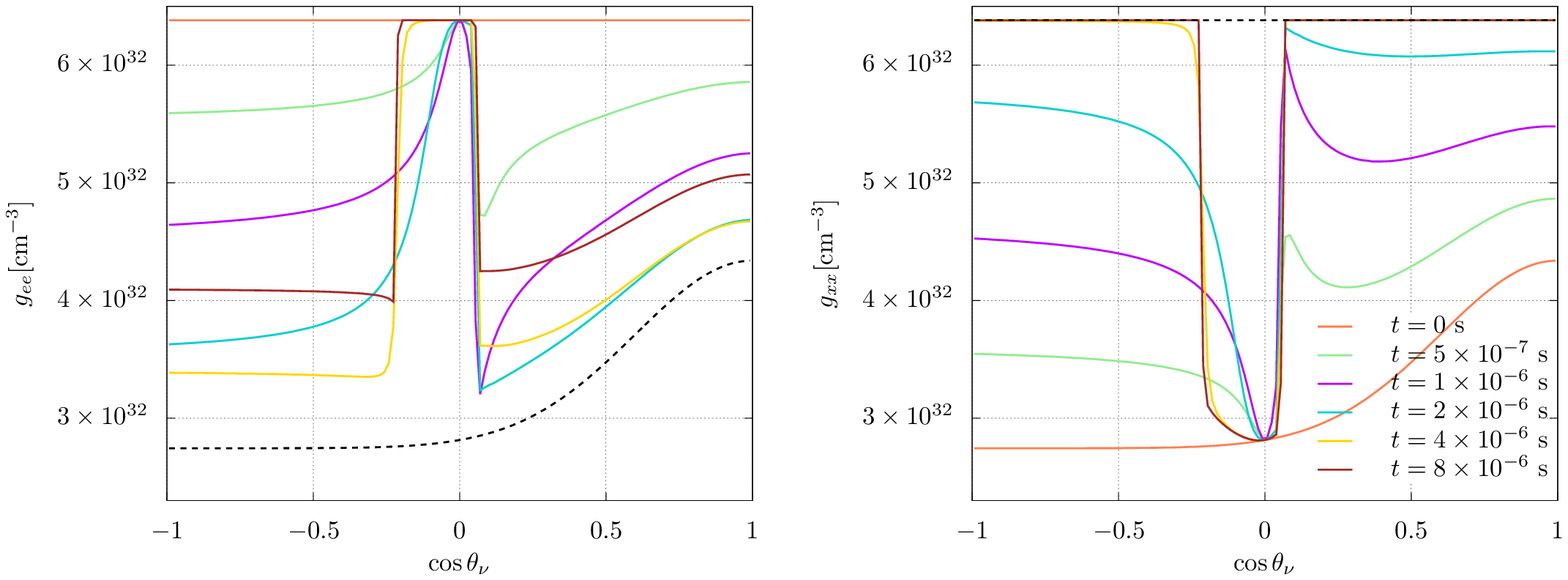}
\caption{The same figures as Figure~\ref{fig:angle_evo} but for the multi-energy case.}
\label{fig:angle_evo_multi}
\end{figure*}

In the late phase ($t\gtrsim4\times10^{-6}$~s), after FFCs are sufficiently attenuated, matter collisions and CFIs drive the evolution.
In top panel of Figure~\ref{fig:number_evo_multi}, we find that $n_{\nu_e}$ increases by emission more slowly than the single-energy case.
This indicates the energy-averaged emission rate is smaller in the multi-energy case, which is consistent with the result in the FFC driven phase.
On the other hand, since reactions for $\nu_x$ and $\bar{\nu}_x$ are neglected, the constant evolution is observed as same as the single-energy case.
As shown by the linear stability analysis, the unstable mode is observed at $t=8\times10^{-6}$~s, but the growth rate $\gamma$ is 
longer than the timescale of interest (see Table~\ref{tab:LSA}).
Considering that the crossings in the $L_e-L_x$ distribution disappears at $t=8\times10^{-6}$~s, this unstable mode is presumed to be CFIs.
Moreover, it should be noted that the asymptotic values of $n_{\nu_x}$ are nearly identical between the single- and multi-energy cases.
This is understood by the evolution of angular distributions in Figure~\ref{fig:angle_evo_multi}.
We find the angular swaps between $\nu_e$ and $\nu_x$ at wider angular range as same as the single-energy case.
As mentioned in Section~\ref{subsec:EAeffects}, FFCs at each angle are determined by the total lepton number and the quantity $P$.
Since adopting the same angular distributions for both cases, the asymptotic states are almost the same.

We also investigate the dependence of initial angular distributions using the ST and PTR models.
In all the models, the initial growth rates $\gamma$ are almost the same as the single-energy case in Table~\ref{tab:LSA}, which again describes the property of FFCs.
The results in the nonlinear phase are shown in Figures~\ref{fig:number_evo_multi} and \ref{fig:LN_evo_multi}. 
In all models, the overall evolution becomes slower than the single-energy cases as same as the PTR-B model, while the features of each model are similar to those of the single-energy case.
Below are some comments on individual models.
The effects of multi-energy treatment in the PTR-C model are similar to that in the PTR-B model.
Specifically, the asymptotic states in the PTR-C model ($t\gtrsim4\times10^{-6}$~s) are almost identical between the single- and multi-energy cases due to the fact that the maximum amount of FFCs is limited.
After the $L_e-L_x$ crossing disappears at $t\sim8\times10^{-6}$~s, CFIs drive the weak flavor conversions.
In the PTR-D model, the $L_e-L_x$ crossings are still observed even at $t=8\times10^{-6}$~s in Figure~\ref{fig:LN_evo_multi} and FFCs survive to the end in the timescale of interest.
As same as the single case of ST model, high angular resolutions are necessary in the multi-energy case.
In the high-resolution results ($N_{\theta_\nu}=512$), the $L_e-L_x$ crossings are found even at $t=8\times10^{-6}$~s and hence the unstable mode at this time may be the weak FFCs.

\section{Summary and discussion} \label{sec:summary}

We perform dynamical simulations of neutrino flavor conversions with neutrino emission and absorption
under the assumption of homogeneous neutrinos.
We adopt physically motivated setups in this study, which is set based on the results of realistic SN simulations \cite{Sumiyoshi:2012za}.
We start with discussing the case with energy-independent reaction rates (Section~\ref{sec:single}), which presents some key features of the interplay between flavor conversion and emission/absorption.
The dynamics is qualitatively different from FFCs without collisions; indeed, the periodic (or pendulum-like) features in flavor conversions disappear.
The result is essentially in line with the case with isoenergetic neutrino-matter scattering \cite{shalgar2021b,sasaki2021,kato2022}.

We also find that the time evolution of flavor conversions with emission/absorption can be divided into two phases, (1)FFC driven phase and (2)collisional one.
The multiple phase appearing in nonlinear regime is accounted for by the disparity of timescales between FFC and collisions.
In the former phase, FFCs are initially more vigorous than the case without collisions due to breaking the symmetry of pendulum motion by collisions.
On the other hand, FFCs are gradually attenuated by matter decoherence.
After the competition of these collision effects, the larger number of $\nu_e$ is converted to $\nu_x$ by FFCs.
In the latter phase, the flavor conversion is driven by CFIs, while the timescale is much longer than FFCs.
$\nu_e$ and $\bar{\nu}_e$, which are less populated than the initial state due to strong FFCs in the earlier phase, are replenished by emission.
The asymptotic states are characterized by matter collisions in this phase.
Although the detailed features depend on the initial condition, we show that the overall trend is similar among all our models. 
It should also be worthy to note that $\nu_x$ emission/absorption do not change our conclusion, unless their reactions become comparable to $\nu_e$ and $\bar{\nu}_e$.

We also find that angular swap between $\nu_e$ ($\bar{\nu}_e$) and $\nu_x$ ($\bar{\nu}_x$) is facilitated by emission/absorption.
In the pure FFC cases, the swap occurs very narrow angular region where ELN crossing occurs nearby. On the other hand, the crossing point can substantially deviate from the initial angular position in the case with emission/absorption, that expands the region where the angular swap occurs.

We extend our discussion in the case with energy-dependent reaction rates.
The initial energy spectrum are set so as to be consistent with realistic SN simulations.
In the low energy regime, the isotropization of neutrinos by emission/absorption is less remarkable, and therefore the lifetime of FFCs can be longer than the single-energy case. As a result, the overall trend becomes similar to the case with the lower reaction rate for monochromatic neutrinos. It should be mentioned that, although the flavor conversion is less vigorous than the single-energy case, the long-lived FFCs compensates for this. In fact, the degree of flavor conversion is almost identical to the signle-energy case, and the angular swap is also observed in a wide angular region. 
We also find that the energy-dependent emission/absorption generates rich energy-dependent features in FFCs similar as scatterings.
In initial condition, the energy spectrum of $\nu_e$ and $\nu_x$ is comparable at $E_\nu\sim25$~MeV, while $\nu_e$ and $\nu_x$ are dominant at the lower and higher energy regions, respectively.
This exhibits that the increase or decrease of each flavor of neutrinos due to flavor conversion becomes qualitatively different between the low and high energy region.

Although the present study does not address all issues in the interplay between flavor conversions and collision term, it provides some valuable insights on quantum kinetic features of neutrinos in CCSN and BNSM environments. One of them is to determine asymptotic states of neutrinos. As shown in our models, FFCs promptly establish a quasi-steady state, which corresponds to, however, just a {\it pseudo} asymptotic state. After FFCs subside, matter collisions or CFIs take over to dictate secure evolution of neutrinos, asymptoting to another state. This indicates that the actual asymptotic state of neutrinos can not be determined solely by either neutrino flavor conversion or neutrino-matter interaction, but by self-consistent treatments of feedback between both of them.

This study also reveals a new possibility that flavor conversion offers a new path to absorb $\nu_x$ in the high energy region.
Since the $\nu_x$ is initially larger than $\nu_e$, flavor conversions facilitate the $\nu_x$ conversion to $\nu_e$, and then it is absorbed through charged current reactions of $\nu_e$, exhibiting the increase of neutrino heating.
It is an intriguing question how the heating can influence on CCSN and BNSM dynamics, although we leave the detailed investigations for future work.

In this work, we impose many simplifications and assumptions such as homogeneous neutrino gas, axial symmetry in momentum space, and two flavor system.
Recent works have shown that the dynamics of flavor conversion is very sensitive to these conditions
(e.g., \cite{abbar2019,martin2020,bhattacharyya2020,sherwood2021b,zaizen2022,Nagakura2022,nagakura2022b,shalgar2022c}).
We also note that the homogeneous approximation suffers from a self-consistent problem in studying the impact of collisions on flavor conversion \cite{lucas2022}.
It should be stressed, however, that some intrinsic features of collision effects can be studied in
our simplified approach.
This offers new insights into roles of flavor conversions on CCSNe/BNSMs dynamics.

\begin{acknowledgements}
We are grateful to Lucas Johns for useful comments and discussions. 
C. K. is supported by JSPS KAKENHI Grant Numbers JP20K14457 and JP22H04577. 
M.Z. is supported by JSPS Grant-in-Aid for JSPS Fellows (No. 22J00440) from the Ministry of Education, Culture, Sports, Science and Technology (MEXT), Japan.
The numerical calculations were carried out on Cray XC50 at Center for Computational Astrophysics, National Astronomical Observatory of Japan.
\end{acknowledgements}

\appendix
\section{Linear stability analysis} \label{ap:LSA}

\begin{figure*}
\centering
\includegraphics[width=\columnwidth]{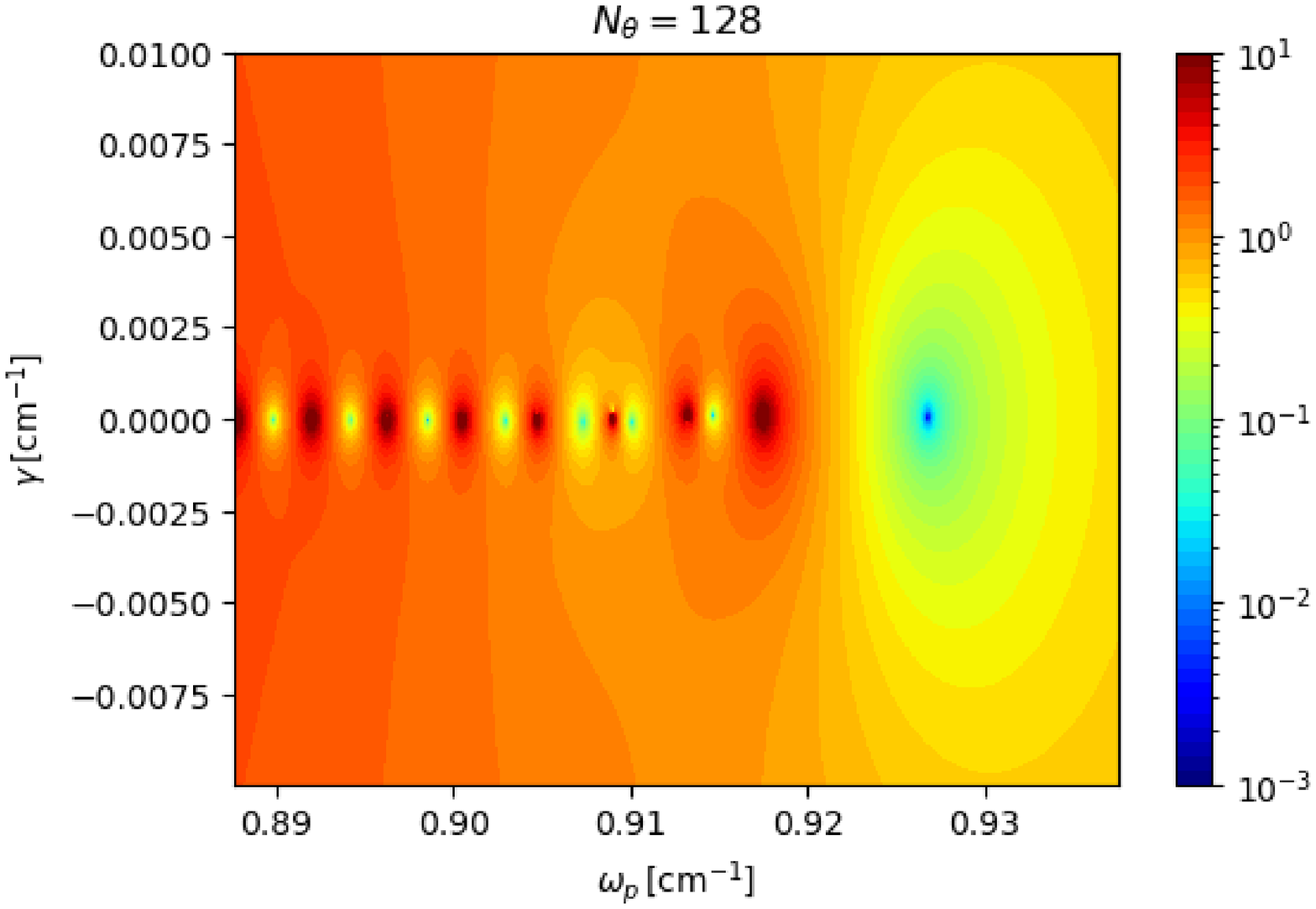}
\includegraphics[width=\columnwidth]{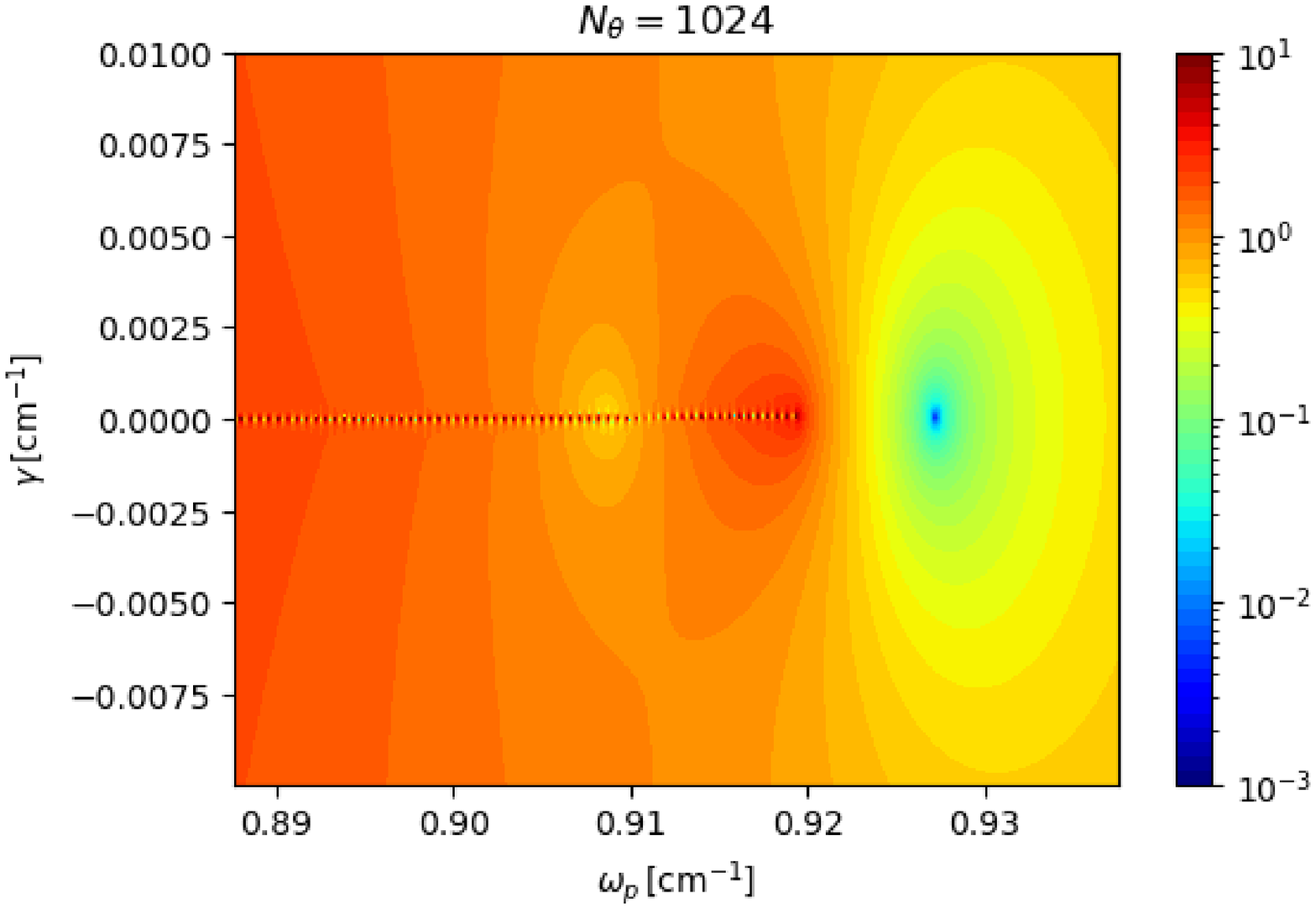}
\caption{Dispersion relation diagrams by the linear stability analysis at $t=4\times10^{-6}$~s in the ST model. Left and right panels show the results of $N_{\theta_\nu}=128$ and 1024, respectively.}
\label{fig:sprious}
\end{figure*}

In the linear stability analysis, we focus on the off-diagonal components $S$ and $\bar{S}$ of neutrino density matrices:
\begin{eqnarray}
\rho = \begin{pmatrix}
\rho_{ee} & S \\
S^\ast & \rho_{xx}
\end{pmatrix},
\bar{\rho} = \begin{pmatrix}
\bar{\rho}_{ee} &\bar{S} \\
\bar{S}^\ast &\bar{\rho}_{xx}
\end{pmatrix}.
\end{eqnarray}
In the flavor-isospin convention, in which negative energy stands antineutrinos as $\bar\rho(E) \equiv -\rho(-E)$,
the governing equation for the off-diagonal component can be written from Eqs.~\ref{eq:rho} and \ref{eq:rhob} as
\begin{eqnarray}
&&\left[i\frac{\partial}{\partial t}+\frac{1}{2}i\left(R_e+R_a\right) \right. \nonumber \\
&&\left. -\sqrt{2}G_F\int dP^\prime \left(1-\cos{\theta_\nu}\cos{\theta_\nu^\prime}\right)\left(\rho_{ee}^\prime-\rho_{xx}^\prime\right) \right] S \nonumber \\
&&= \sqrt{2}G_F \left(\rho_{xx}-\rho_{ee}\right)
\int dP^\prime \left(1-\cos{\theta_\nu}\cos{\theta_\nu^\prime}\right)S^\prime, \ \ \ \ \ \ 
\end{eqnarray}
with $dP^\prime = E_\nu^{\prime2}dE_\nu^\prime d\cos{\theta_\nu^\prime d\phi_\nu^\prime}/(2\pi)^3$.
In the linear stability analysis, we assume that the off-diagonal components are enough small to ignore the mode coupling with the diagonal components.
Then, for the plane wave solution $S\propto Q\exp(-i\Omega t)$ in a homogeneous mode, we obtain
\begin{eqnarray}
&&\left[\Omega-\left(\Lambda_{0}-\Lambda_{1}\cos\theta_{\nu}\right)+i\Gamma\right]Q \notag\\ 
&&\,\,\,\,\,= - \left(a_0-a_1\cos\theta_{\nu}\right)\left(\rho_{ee}-\rho_{xx}\right), \label{eq:kspace}
\end{eqnarray}
where
\begin{eqnarray}
\Gamma &\equiv& \frac{1}{2}\left[R_e(E_{\nu})+R_a(E_{\nu})\right], \\
\Lambda_{0} &\equiv& \sqrt{2}G_F\int dP \left(\rho_{ee}-\rho_{xx}\right), \\
\Lambda_{1} &\equiv& \sqrt{2}G_F\int dP \cos{\theta_\nu}\left(\rho_{ee}-\rho_{xx}\right), \\
a_0 &\equiv& \sqrt{2}G_F\int dP\, Q, \label{def:a}\\
a_1 &\equiv& \sqrt{2}G_F\int dP \cos{\theta_\nu}\, Q. \label{def:ak}
\end{eqnarray}
By solving Eq.~\ref{eq:kspace}, we derive the eigenfunction $Q$ as
\begin{eqnarray}
    Q = - \frac{\left(a_0 - a_1 \cos\theta_{\nu}\right)\left(\rho_{ee}-\rho_{xx}\right)}{\Omega-\left(\Lambda_{0}-\Lambda_{1}\cos\theta_{\nu}\right)+i\Gamma},
\end{eqnarray}
and also describe $a_0$ and $a_1$ as
\begin{eqnarray}
    a_0 &=& -\sqrt{2}G_F\int dP \frac{\left(a_0 - a_1 \cos\theta_{\nu}\right)\left(\rho_{ee}-\rho_{xx}\right)}{\Omega-\left(\Lambda_{0}-\Lambda_{1}\cos\theta_{\nu}\right)+i\Gamma}, \\
    a_1 &=& -\sqrt{2}G_F\int dP \cos\theta_{\nu} \frac{\left(a_0 - a_1 \cos\theta_{\nu}\right) \left(\rho_{ee}-\rho_{xx}\right)}{\Omega-\left(\Lambda_{0}-\Lambda_{1}\cos\theta_{\nu}\right)+i\Gamma}. \notag\\
\end{eqnarray}
The matrix expression of above two equations is
\begin{eqnarray}
    \begin{pmatrix}
        a_0 \\ a_1
    \end{pmatrix}
    = \begin{pmatrix}
        -I_0 & I_1 \\ -I_1 & I_2
    \end{pmatrix}
    \begin{pmatrix}
        a_0 \\ a_1
    \end{pmatrix},
\end{eqnarray}
where
\begin{eqnarray}
    I_n = \sqrt{2}G_F\int dP \cos^n\theta_{\nu}\frac{\left(\rho_{ee}-\rho_{xx}\right)}{\Omega-\left(\Lambda_{0}-\Lambda_{1}\cos\theta_{\nu}\right)+i\Gamma}, \notag\\ \label{eq:In_DR}
\end{eqnarray}
and the nontrivial solutions for $a_0$ and $a_1$ exist when the following condition is satisfied
\begin{eqnarray}
    D(\Omega) \equiv (1+I_0)(1-I_2) + I_1^2 = 0. \label{def:D}
\end{eqnarray}
We derive a dispersion relation $\Omega\equiv\omega_p+i\gamma$ for the homogeneous mode as the solution of Eq.~\ref{def:D}.
In Figures~\ref{DR_modelB}, \ref{fig:modelB_wocol}, \ref{DR_initial_all} and \ref{fig:sprious}, we plot $\left|D\right|$ on the $\omega_p-\gamma$ plane for the root-finding. The values of $|D|$ are described by the color bar and the combinations of $\omega_p$ and $\gamma$, where $|D|\sim0$ (blue regions), are the dispersion relations of the systems. 
The $\gamma>0$ ($\gamma\leqq0$) solutions correspond to growing (decaying) modes.

The origin of CFI modes is the collision rates $\Gamma$ in the denominator of Eq.~\ref{eq:In_DR}. If we find the $\gamma>0$ solutions, they are possible CFI modes. In the collisionless case, or $\Gamma=0$, on the other hand, Eq.~\ref{eq:kspace} becomes the equation for the pure FFCs in the homogeneous case. By solving Eq.~\ref{def:D} with $\Gamma=0$, we can determine whether a system is stable or unstable for FFC modes. Moreover, the analysis of $\Gamma=0$ tells us important information for CFI modes. Since $\Gamma$ is always real, it changes not the real part of $\Omega$ ($\omega_p$) but the imaginary part ($\gamma$). Hence if the $\gamma>0$ modes in the $\Gamma\ne0$ analysis become stable ones in the $\Gamma=0$ analysis at the same $\omega_p$, we confirm that they are CFI modes.

It should be noted that we can see the presence of poles (or branch cuts) leading to spurious modes in the denominator in Eq.\,\ref{eq:In_DR} \cite{morinaga2018}.
For $\Lambda_{0,1}>0$, the branch cut spans from $\Lambda_0-\Lambda_1$ to $\Lambda_0+\Lambda_1$ on the axis of $\gamma=i\Gamma(E_{\nu})$ and is replaced with spurious modes by discretizing the angular integration in Eq.\,\ref{eq:In_DR}.
For example, Figure~\ref{fig:sprious} shows the results of linear stability analysis for the ST model at $t=4\times10^{-6}$~s.
Left and right panels show the cases with $N_{\theta_\nu}=128$ and 1024, respectively. Spurious modes appear at different positions in the dispersion relation diagram depending on angular resolution, whereas regardless of angular resolution, the actual solution appear in the same positions.
Following this strategy, we select the actual solution of the dispersion relations.
For this example, the actual solution is found at $\omega_p\sim 0.925$.

\bibliography{MCEA}

\end{document}